\documentclass[preprint,  review, 12pt]{elsarticle}
    
\usepackage{amssymb}
\usepackage{amsmath}
\usepackage{soul}
\usepackage[dvipsnames]{xcolor}
\usepackage[left]{lineno}

\journal{Journal of Molecular Liquids}

\begin{document}

\begin{frontmatter}



\title{Dielectric spectroscopy of a ferroelectric nematic liquid crystal and the effect of the sample thickness}


\author[inst1]{Aitor Erkoreka}

\affiliation[inst1]{organization={Department of Physics, Faculty of Science and Technology, University of the Basque Country UPV/EHU},
            city={Bilbao},
            country={Spain}}
\author[inst1]{Josu Martinez-Perdiguero\corref{cor1}}
\ead{jesus.martinez@ehu.eus}
\cortext[cor1]{Corresponding author}

\author[inst2,inst3] {Richard J. Mandle}

\affiliation[inst2]{organization={School of Physics and Astronomy, University of Leeds},
            city={Leeds},
            country={UK}}

\affiliation[inst3]{organization={School of Chemistry, University of Leeds},
            city={Leeds},
            country={UK}}
            
\author[inst4]{Alenka Mertelj}
\author[inst4]{Nerea Sebastián}

\affiliation[inst4]{organization={Jožef Stefan Institute},
            city={Ljubljana},
            country={Slovenia}}

\let\today\relax
\makeatletter
\def\ps@pprintTitle{%
    \let\@oddhead\@empty
    \let\@evenhead\@empty
    \def\@oddfoot{\footnotesize\itshape
         {} \hfill\today}%
    \let\@evenfoot\@oddfoot
    }
\makeatother

\begin{abstract}
The recently discovered ferroelectric nematic liquid crystals have been reported to exhibit very large dielectric permittivity values. Here, we report a systematic investigation of the dielectric behavior of a prototypical ferroelectric nematogen by varying the thickness of the parallel capacitor measuring cell. While in the non-polar high temperature nematic phase results show only slight differences due to slight variations of the alignment, the measured permittivity values in the ferroelectric nematic phase show a linear dependence on the cell thickness. It is also shown that the characteristic relaxation frequency decreases inversely proportionally to the thickness. The results are discussed in terms of three different available models based on different underlying mechanisms, accounting for cancellation of the probe electric fields by polarization reorientation or by ionic charges, or based on a recently proposed continuous phenomenological model.
\end{abstract}


\end{frontmatter}

\section{Introduction}
\label{sec:introduction}
Among the whole variety of liquid crystal (LC) mesophases discovered since the first mesogenic materials were found, the classical non-polar nematic phase (N) is the one monopolizing their widespread application, such as in display technologies. This uniaxial phase, in which molecules show long-range orientational order but no long-range translational order, exhibits a unique combination of anisotropic physical properties and fluidity. The director (i.e. the direction of the average molecular orientation) can be easily reoriented via the application of small electric fields, resulting in an interesting range of exploitable electro-optic behavior. Even if the constituent molecules possess an electric dipole, the N phase is non-polar, since at any moment there are just as many molecules pointing in one particular direction as in the opposite. Interestingly, and despite its prediction at the beginning of the past century \cite{debye, born}, a polar version of the N phase, in which such inversion symmetry is broken, has been elusive until 2017, when two different materials showing nematic phases with ferroelectric properties were found \cite{mandle_nematic_2017, nishikawa18}. The polarity of the phase is evidenced both by the very high spontaneous polarization values in the $\mu\mathrm{C}/\mathrm{cm}^2$ range \cite{nishikawa18, chen_first-principles_2020, chen_ideal_2022, brown_multiple_2021} and by the large values of the nonlinear optical coefficients \cite{folcia_ferroelectric_2022, li_how_2021, li_development_2021}.

Among the outstanding properties of the ferroelectric nematic phase (N$_{\mathrm{F}}$) are also the reported values of dielectric permittivity, of the order of $10^3-10^5$ (e.g. \cite{li_development_2021, manabe_ferroelectric_2021, yadav_polar_2022, nishikawa_new_2021, nishikawa_anisotropic_2022}). Such range, spanning up to two orders of magnitude, raises the question of how measurement conditions influence the results and whether reported values can be fully attributed to the material properties \cite{brown_multiple_2021, clark_dielectric_2022, vaupotic_dielectric_2022}. These measurements correspond to various experimental conditions in terms of sample thicknesses, anchoring conditions and electrode resistivities. The magnitude of polar correlations in these materials makes dielectric spectroscopy a highly relevant technique for their investigation, not only in the case of the N$_{\mathrm{F}}$ phase, but the subsequently described antiferroelectric phase (M2/N$_{\mathrm{S}}$/SmZ$_A$ phase) observed for some representative materials in between the N and the N$_{\mathrm{F}}$ phase \cite{brown_multiple_2021}, as well as for the reported chiral counterpart of N$_{\mathrm{F}}$ \cite{nishikawa_new_2021, zhao_spontaneous_2021, chiral_tunable,ortega_ferroelectric_2022, POCIECHA2022119532} or the lower temperature ferroelectric SmA phases, for which polarization lies along the layer normal \cite{chen_observation_2022, kikuchi_fluid_2022}. Investigating the dielectric response of these systems is not only important to achieve a fundamental understanding of the physical processes leading to these phases, but also for practical applications involving these materials such as faster and lower-consumption displays, improved electro-optic modulators, high-performance electronics, memory devices, high-density capacitors, etc.

Clark et al. have recently proposed that what is being measured in these experiments is not the bulk sample but the high capacitance of the insulating interfacial layers of nanoscale thickness that bound the sample in typical measurement cells \cite{clark_dielectric_2022, shen_effective_2011}, thus leading to an erroneous interpretation of the experimental results. In fact, the N$_{\mathrm{F}}$ phase is expected to show a Goldstone mode (phason), by analogy with other ferroelectric liquid crystals \cite{kremer_broadband_2003}. This mode would correspond to a reorientation of the polarization vector which, coupled with a high fluidity, would enable the charging of the insulating layers and screen the electric field in the bulk sample (Clark et al. call it polarization-external capacitance Goldstone reorientation mode, or PCG for short \cite{clark_dielectric_2022}). Since this is an interfacial effect, the model predicts a relaxation process whose characteristics depend on the sample thickness $d$. In another recent article, Vaupoti\u{c} et al. \cite{vaupotic_dielectric_2022} interpret experimental observations through the development of a continuous phenomenological model for the N$_{\mathrm{F}}$ phase, in which director and polarization fluctuations are coupled, giving rise to two polar modes (phason and amplitudon modes). Despite the authors subsequently neglecting it, their theoretical model predicts an intrinsic dependence on $d$. Finally, it should be noted that electrode polarization (EP) is a well-studied phenomenon known to affect the dielectric spectra of conductive samples such as solid-state electrolytes, aqueous solutions, ionic liquids, biological systems, etc \cite{kremer_broadband_2003}, leading to extremely high measured values of the real and imaginary components of the dielectric permittivity, especially at low frequencies. As several models rigorously prove, it turns out that the dynamics of the ions depends on the sample thickness $d$ \cite{chassagne_compensating_2016}. The spectra of LCs can also be influenced by this effect, since they always contain ionic impurities. In the case of ferroelectric nematic materials, ionic impurities play an important role in the observed structures under confinement \cite{basnet_soliton_2022, patterning} and should not be neglected in the interpretation of dielectric measurements. In any case, the three frameworks presented above predict that the experimental dielectric spectra depend on the thickness of the parallel-plate capacitor used in the measurements.

With the aim of gaining insight into the origin of the large values of the measured dielectric permittivity, in this paper we perform a systematic set of broadband dielectric spectroscopy experiments for RM734, one of the representative N$_{\mathrm{F}}$ materials, varying the measuring cell thickness. By identifying characteristic features of the spectra, we interpret the results in the view of the above-mentioned possible mechanisms.

\section{Materials and Methods}
\label{sec:materials}
\subsection{Material}
\begin{figure}[ht!]
\begin{center}
\includegraphics[width=80 mm]{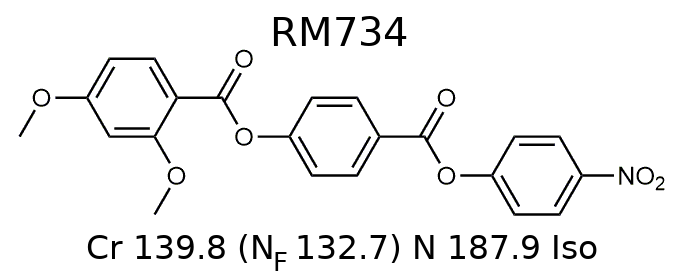}
\end{center}
\caption{\label{fig:molecule} Chemical structure of RM734 and its transition temperatures on cooling.}
\end{figure}

Dielectric investigations were performed for RM734 (4-((4-
nitrophenoxy)\-carbonyl)phenyl-2,4-dimethoxybenzoate), a representative ferroelectric nematic material, which was synthesized via literature methods \cite{mandle_rational_2017}. Its molecular structure, phase sequence and transition temperatures are shown in Fig. \ref{fig:molecule}. On cooling, RM734 exhibits a direct N-N$_{\mathrm{F}}$ phase transition. The N$_{\mathrm{F}}$ phase can be reached in a supercooled state with cooling rates even lower than $0.25^{\circ}\mathrm{C}/\mathrm{min}$ and is stable, in almost all runs, well below $80^{\circ}\mathrm{C}$. RM734 molecules have a large dipole moment ($\sim$11.4 D \cite{mertelj_splay_2018}) directed at a moderate angle (20$^{\circ}$) from the molecular axis.

\subsection{Broadband dielectric spectroscopy}
Measurements of the complex dielectric permittivity $\varepsilon(f) = \varepsilon'(f)-i\varepsilon''(f)$ were carried out from 1 Hz up to 3 MHz with an Alpha-A impedance analyzer from Novocontrol Technologies GmbH.  The material was placed in the N phase ($\sim 160^{\circ}$C) between two circular gold-plated brass electrodes 5 mm in diameter acting as a parallel-plate capacitor. For improved repeatability, the same pair of electrodes was thoroughly cleaned and used in all measurements. The separation between electrodes was fixed by spherical silica spacers of different diameters. The low resistivity of gold electrodes results in a cut-off frequency of the measuring circuit which is much higher than the frequency range of the experiments (see for instance \cite{cut_off}). The sample was placed at the end of a modified HP16091A fixture, and the temperature of the sample was controlled with a Quatro cryostat also from Novocontrol. The stray capacitance of the measurement circuit was carefully taken into account.

It is well-known that the use of alignment layers can give rise to charge accumulation and additional undesired effects \cite{brown_multiple_2021, vaupotic_dielectric_2022, shuichi_murakami_electrode_1997, 5cb} and, consequently, the measurements here were performed with the bare metal electrodes and no surface aligning layers. In order to accurately compare the results for the different samples, the real thickness of each measuring cell was calibrated by normalizing the dielectric permittivity value in the isotropic phase with that obtained in a 7 $\mu$m-thick sandwich commercial cell (EHC Co. Ltd, Japan). The deviation from the nominal values was small, which confirms the good performance of the silica spacers. Obtained thicknesses range between 7 and 104 $\mu$m. The measured complex dielectric permittivity was thus calculated by dividing the measured capacitance by the theoretical geometric capacitance of our measuring parallel plate capacitor for each given thickness. To minimize the time spent at high temperatures, all samples were heated up to $190^{\circ}$C and then cooled at $5^{\circ}\mathrm{C}/\mathrm{min}$ down to $160^{\circ}$C (nematic phase) while performing fast characterization measurements at a few given frequencies. All measurements were then performed on cooling from $160^{\circ}$C at $0.25^{\circ}\mathrm{C}/\mathrm{min}$ with $V_{\mathrm{osc}}=0.03$ $V_\mathrm{rms}$. The measuring voltage was kept constant regardless of the sample thickness. The effect of increasing probe voltages was also analyzed as will be shown below. Finally, for the sake of completeness, the dielectric response of the sample under the application of a DC bias field was studied.

\begin{figure}[b!]
\begin{center}
\includegraphics[width=79 mm]{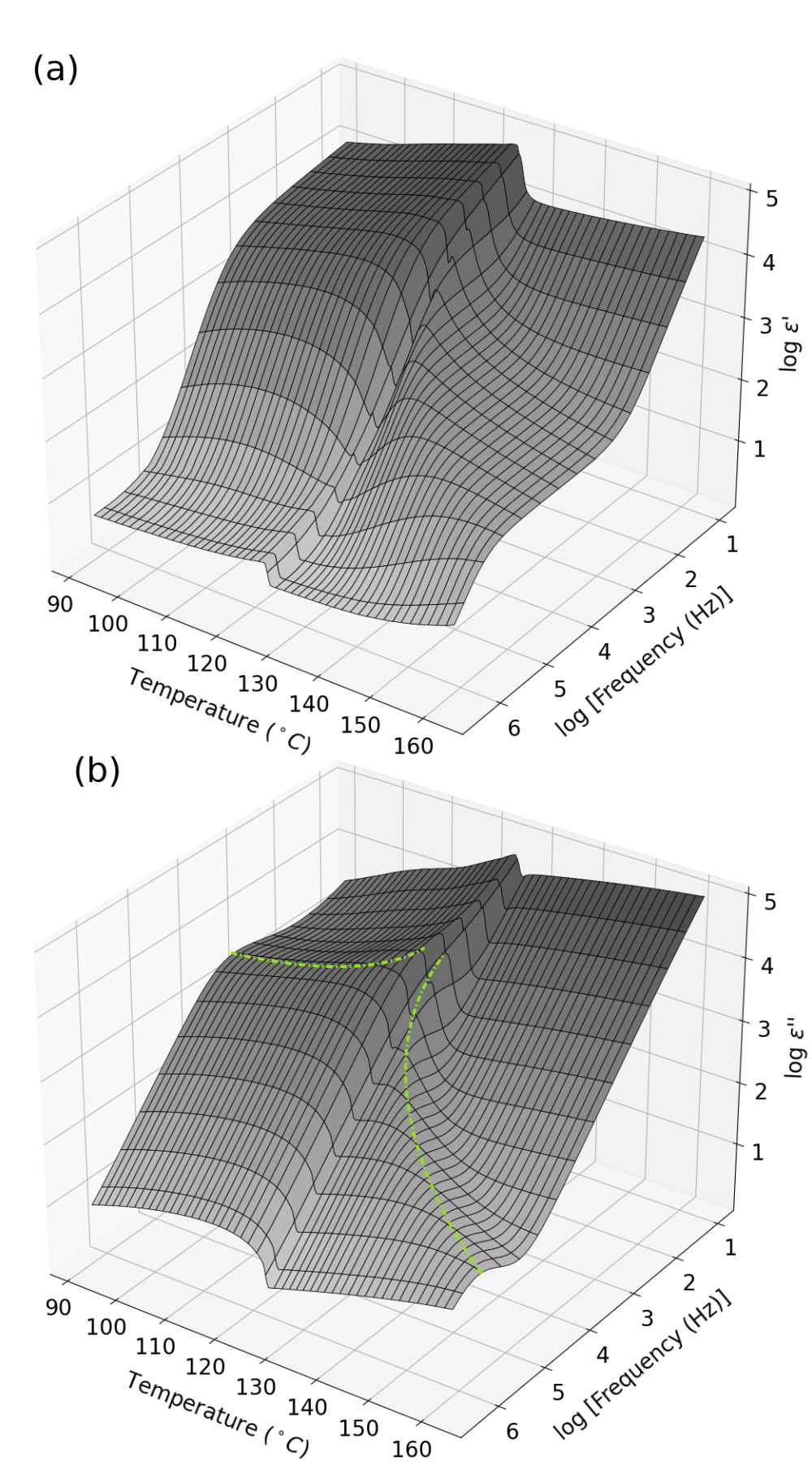}
\end{center}
\caption{\label{fig:3dplot} Three-dimensional plot of the real (a) and imaginary (b) components of the complex permittivity vs. temperature and frequency for a $d=26\:\mu$m thick cell. The dashed line in (b) highlights the temperature dependence of the maximum absorption frequency.}
\end{figure}

\section{Results}
\label{sec:results}
As an example of the performed measurements, Fig. \ref{fig:3dplot} shows the permittivity components $\varepsilon'$ and $\varepsilon''$ as a function of frequency and temperature for a $d=26\:\mu$m thick cell. The measured values are very large, of the order of $10^4$ at low frequencies, both in the higher temperature non-polar nematic phase and in the N$_{\mathrm{F}}$ phase.

\begin{figure}[b!]
\begin{center}
\includegraphics[width=70 mm]{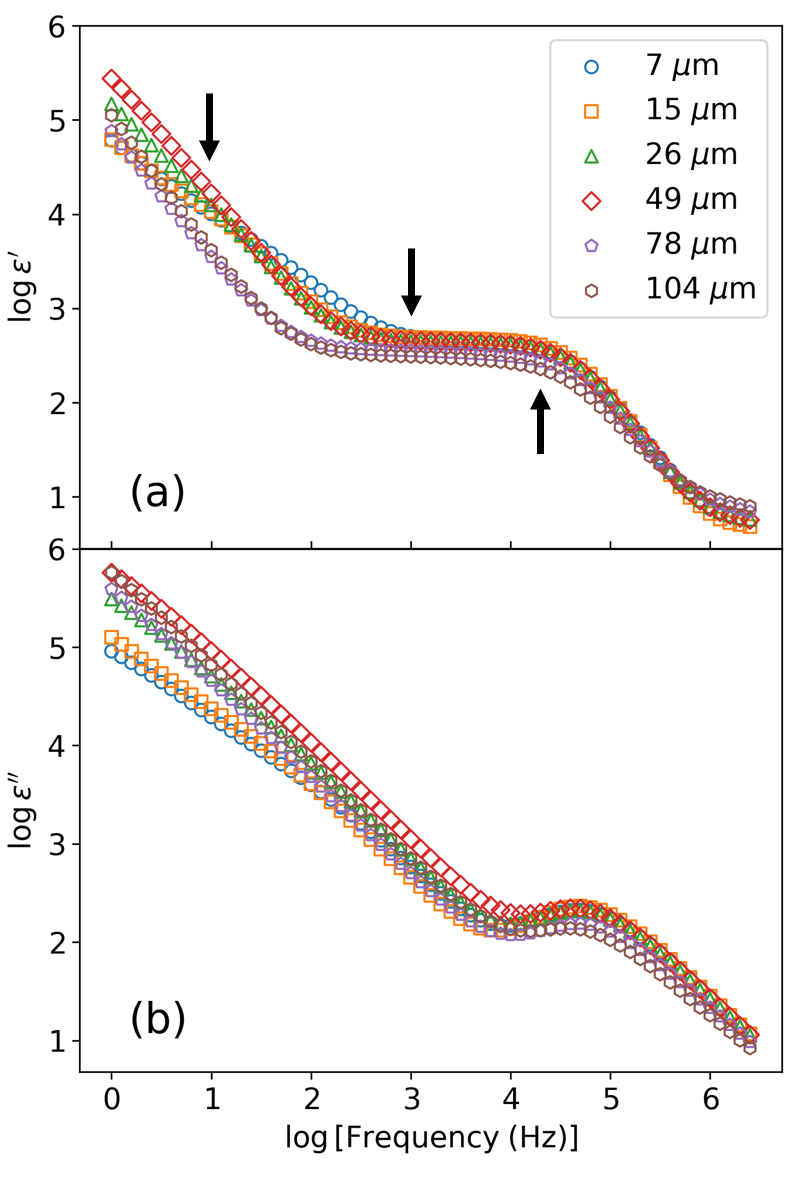}
\end{center}
\caption{\label{fig:nematic140} Spectrum of the real (a) and imaginary (b) components of the complex permittivity for the different electrode separations at 140$^{\circ}$C (N phase). Black arrows indicate the frequencies shown in Fig. \ref{fig:tempDep}.}
\end{figure}

We first focus on the non-polar N phase. A comparison of the recorded spectra at 140$^{\circ}$C is shown in Fig. \ref{fig:nematic140} for the different cell thicknesses. The recorded spectra are characterized by a relaxation process of frequency around $10^5$ Hz and strength around 300. Due to its strength, this relaxation process can only be associated with a collective fluctuation of dipole moments and has been shown to strongly soften when approaching the N-N$_{\mathrm{F}}$ transition, with a rapidly decreasing frequency and diverging amplitude \cite{brown_multiple_2021, PhysRevLett.124.037801, mandle_molecular_2021}. The slight differences in strength evident in Fig. \ref{fig:nematic140} can be attributed to slight differences in alignment as will be shown below. Of interest in the present context is the sharp increase of permittivity values below 100 Hz. Such increase is typical of EP processes and is well known for measurements of non-polar nematic materials such as 5CB \cite{shuichi_murakami_electrode_1997, murakami_dielectric_1996, 5cb_prakash}. The characteristic frequency of this process depends on the diffusive motion of impurity ions. In the N phase, it occurs at low enough frequencies so that the actual spectra of the material can be discerned.

\begin{figure}[t!]
\begin{center}
\includegraphics[width=70 mm]{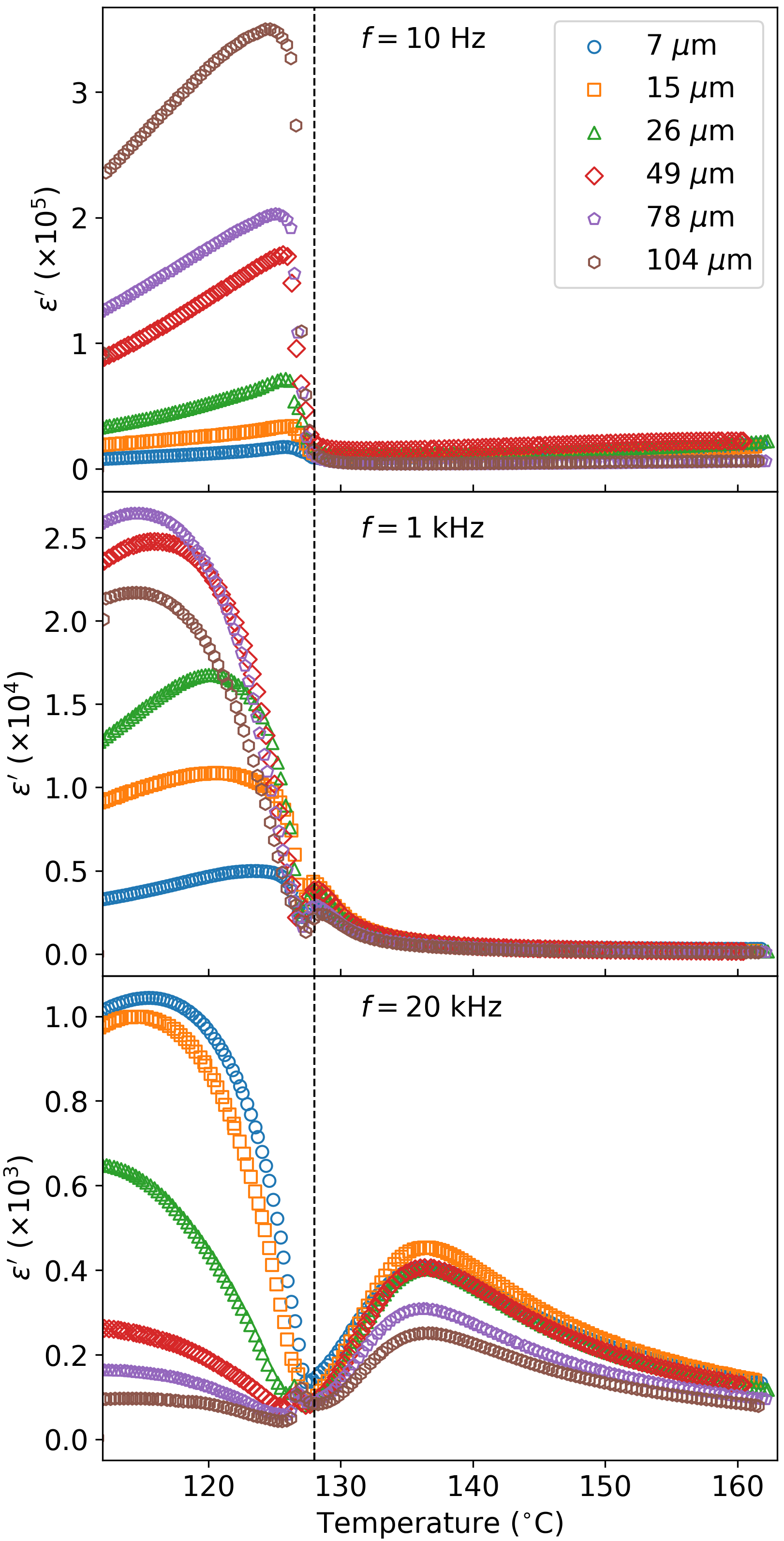}
\end{center}
\caption{\label{fig:tempDep} Temperature dependence of the measured dielectric permittivity for three different frequencies (10 Hz, 1 kHz and 20 kHz) as measured on cooling for different electrode separations.}
\end{figure}

The temperature dependence of the recorded permittivity values across the N-N$_\mathrm{F}$ phase transition at 10 Hz, 1 kHz and 20 kHz for the different cells is plotted in Fig. \ref{fig:tempDep}. As it can be seen, the transition temperature is around $128^{\circ}$C, below that initially reported in Fig. 1 obtained via DSC. This difference could be attributed to a slight increase of impurities over time, which can have an effect on the transition temperatures due to internal field screening effects \cite{patterning}. The comparison in the N phase at 20 kHz corresponds to the range of frequencies at which EP does not contribute anymore. Although the temperature dependence tendency is the same, it is clearly observed how the permittivity value decreases for thicker cells. It should be recalled that no alignment layers have been employed and, thus, such differences can be easily attributed to differences in alignment. The increase of $\varepsilon'$ at the isotropic to N phase transition (Fig. S1) indicates that in the gold electrodes, RM734 tends to align homeotropically, i.e. with the director perpendicular to the electrode surface. However, it is reasonable to assume that the alignment is not perfect and that the increase of $d$ results in the decrease of surface-induced elastic forces in the cell center that will preserve the homeotropic alignment.

\begin{figure}[t!]
\begin{center}
\includegraphics[width=70 mm]{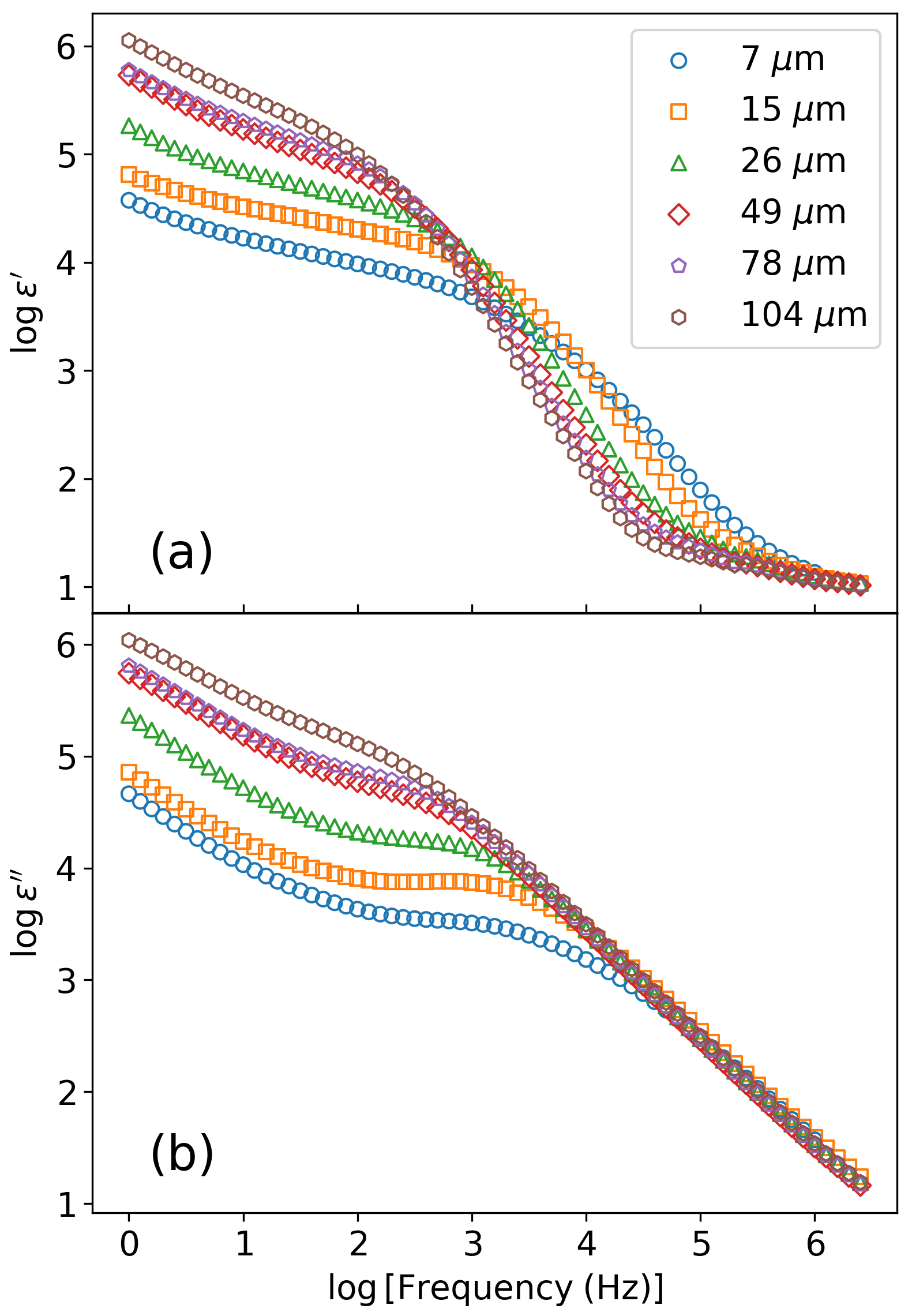}
\end{center}
\caption{\label{fig:measured_curves} Spectrum of the real (a) and imaginary (b) components of the complex permittivity for the different electrode separations at $125^{\circ}$C (N$_{\mathrm{F}}$ phase).}
\end{figure}

At the transition to the polar nematic phase, the large polarization values imply that it is reasonable to assume that, in order to minimize the depolarization field, the phase transition is accompanied by a change of director orientation from homeotropic to planar \cite{nishikawa18, nishikawa_anisotropic_2022, PhysRevLett.124.037801, Saha_Nepal, caimi}. This means that in the N$_\mathrm{F}$ phase the director lies parallel to the cell surfaces. This is further supported by the very large electric fields that have been reported to be required to induce the transition from planar to homeotropic configuration in the N$_\mathrm{F}$ phase \cite{Saha_Nepal}. In terms of the measured permittivity as a function of cell thickness, very different behavior is observed in the N$_\mathrm{F}$ phase. At the same frequency of 20 kHz, the recorded permittivity greatly increases for decreasing cell thickness (from values around 100 at $d=104\:\mu$m to values slightly higher than 1000 at $d=7\:\mu$m). This tendency is inverted at lower frequencies, for which the thicker the measuring cell, the larger the value of the recorded permittivity (top graph of Fig. \ref{fig:tempDep}). It should be noted here that Manabe et al. \cite{manabe_ferroelectric_2021} reported a similar behavior for a material showing a direct isotropic to N$_\mathrm{F}$ transition, for which they performed measurements at two different thicknesses (110 and 10 $\mu$m). However, it should be pointed out that, in their case, the measuring cell not only varied in thickness but also in area, electrode material and aligning agent. Their reported values correspond to 1 kHz, in the range at which there is no clear correlation for RM734 (middle graph of Fig. \ref{fig:tempDep}). This can be easily clarified by inspecting the spectra in the N$_\mathrm{F}$ phase, which are shown in Fig. \ref{fig:measured_curves} at $125^{\circ}$C. For completeness, the directly measured complex impedance is available in Fig. S2.

\begin{figure}[t!]
\begin{center}
\includegraphics[width=70 mm]{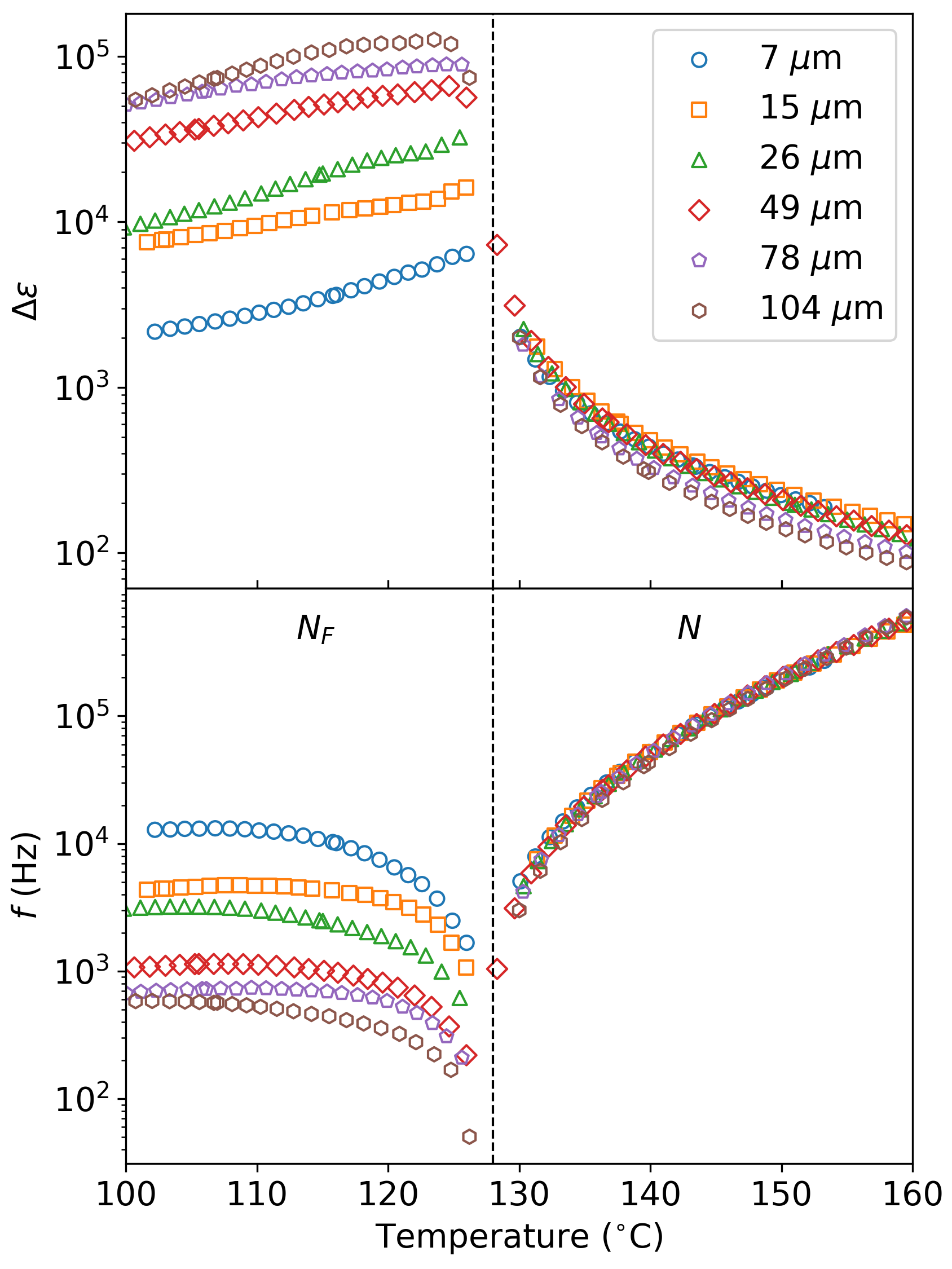}
\end{center}
\caption{\label{fig:Fits} Strength ($\Delta\varepsilon$) and frequency of maximum absorption of the main relaxation mode as obtained from fitting the data to an HN relaxation for the different cell thicknesses. An extra very low-frequency mode (not represented here) was employed to account for the increase of $\varepsilon$ at low frequencies observed in Fig. \ref{fig:measured_curves}.}
\end{figure}

In the N$_\mathrm{F}$ phase, the spectra are characterized by a broad relaxation process with a characteristic frequency in the 1-10 kHz range. It should be noted that additional low frequency effects can be observed. Focusing on the real part $\varepsilon'$, at low frequencies, the value increases with increasing thickness. Around $\sim 1$ kHz there is a crossover frequency at which that trend is inverted. Finally above $\sim 1$ MHz  the values for all thicknesses agree. This behavior can be qualitatively explained as a thickness-dependent relaxation: the amplitude grows with $d$ while the frequency decreases. This becomes evident when inspecting the results from fitting the recorded spectra to a Havriliak-Negami (HN) relaxation (see Eq. S1), as shown in Fig. \ref{fig:Fits} (HN parameters and fitting examples are available in Table S1 and Figure S3 respectively). It is seen that the frequency of the main relaxation mode in the polar phase increases with decreasing cell thickness, with a difference of up to a decade between the thinnest $d=7\:\mu$m cell and the thickest $d=104\:\mu$m cell. Fitting results in the N phase additionally show how the dielectric spectra for all the inspected thicknesses are fully equivalent, where the softening behavior is clearly observed. It should be mentioned that the growth of polar correlations in the N phase has been shown to be accompanied by a strong softening of the splay elastic constant \cite{PhysRevLett.124.037801}. This softening behavior is characteristic of the N-N$_\mathrm{F}$ transition, and both dielectric and nematic elasticity softening are absent for a RM734 homologue that does not exhibit the N$_\mathrm{F}$ phase \cite{mandle_molecular_2021}.

\begin{figure}[t!]
\begin{center}
\includegraphics[width=70 mm]{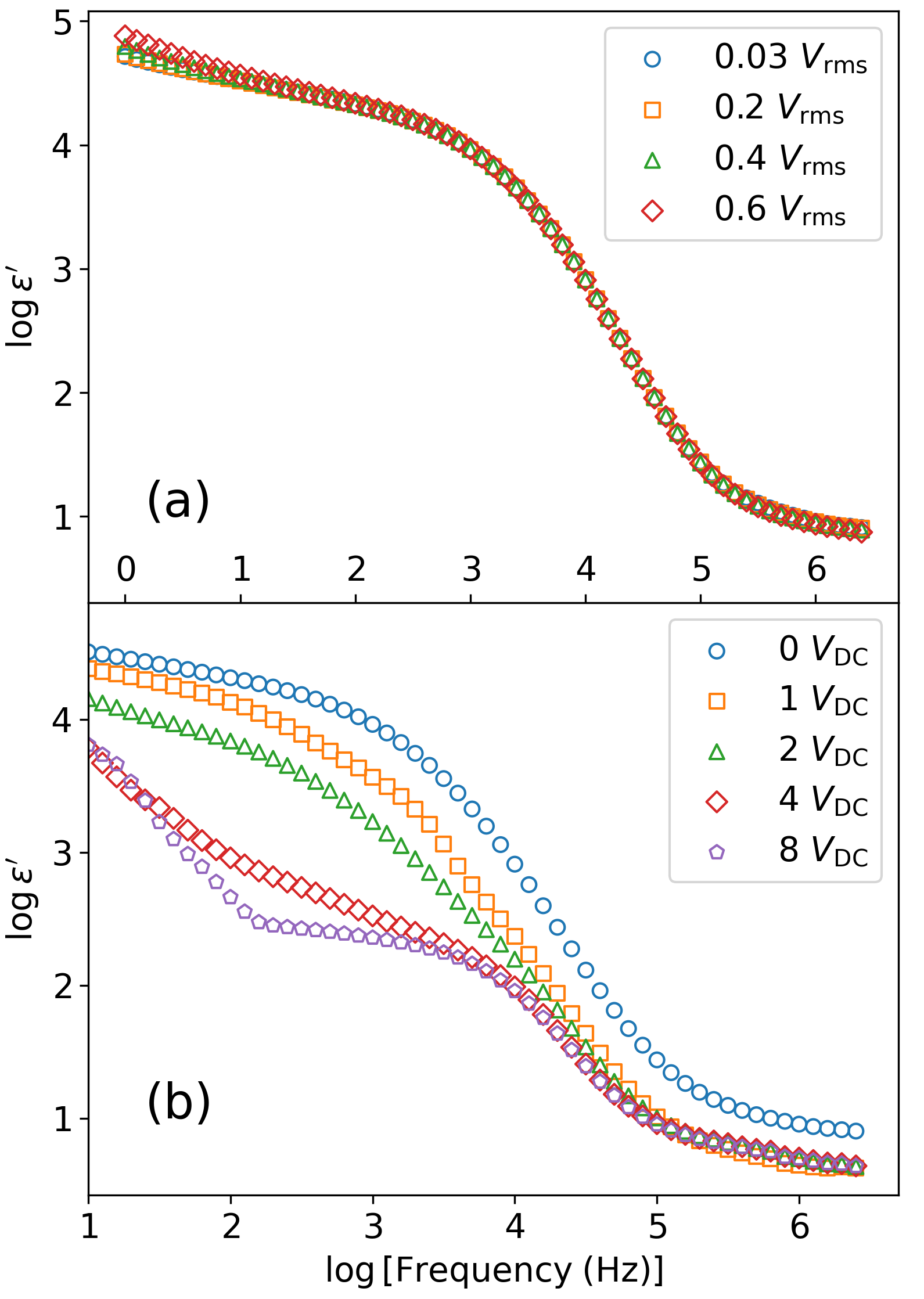}
\end{center}
\caption{\label{fig:ACDC} Spectrum of the real component of the complex permittivity at $120^{\circ}$C} for a cell of 25 $\mu$m at different probe (AC) (a) and DC-bias (b) voltages.
\end{figure}

As mentioned in Section \ref{sec:materials}, to avoid field-induced reorientations in the N phase due to low splay elastic constants \cite{mertelj_splay_2018, PhysRevLett.124.037801, mandle_molecular_2021} all measurements were carried out with a low oscillator voltage of 0.03 V$_\mathrm{rms}$. To test the influence of this voltage in the measurements acquired in the N$_\mathrm{F}$ phase, we performed a series of frequency sweeps with oscillator values up to 0.6 V$_\mathrm{rms}$ in a 25 $\mu$m thick cell, where the last value corresponds to the same oscillator level reported by Vaupotič et al. \cite{vaupotic_dielectric_2022}. The results, shown in Fig. \ref{fig:ACDC}a, only vary in the low-frequency regime where a small increase of both $\varepsilon$ is observed. For completeness, we also studied the effect of the application of DC fields in this same cell thickness (Fig. \ref{fig:ACDC}b). The effect is much larger in this case, with a strong reduction of the $\varepsilon'$ value at low frequencies, due to the probable shift to lower frequencies of a relaxation process, revealing a second process at moderate DC fields. For the 8 V DC curve (larger fields only resulted in noisier data and no quantitative changes were measured), data can be fitted to a relaxation process with an amplitude of around 200 in the 10 kHz frequency range. It is interesting to note that there is a decrease of  $\varepsilon'$ at high frequencies likely due to the suppression of a high-frequency molecular mode. Finally, it is worth mentioning that after the application of the DC voltage, the initial signal was fully recovered, indicating no degradation of the material.

\section{Discussion}
The aforementioned PCG \cite{clark_dielectric_2022} and EP \cite{chassagne_compensating_2016} models, as well as the continuous phenomenological model (CPM hereafter) by Vaupotič et al. \cite{vaupotic_dielectric_2022}, introduce dependence of the measured spectra on $d$. In the present section, we will analyze our data in this context.

\subsection{Analysis in terms of PCG and CPM}
\begin{figure}[t!]
\begin{center}
\includegraphics[width=75 mm]{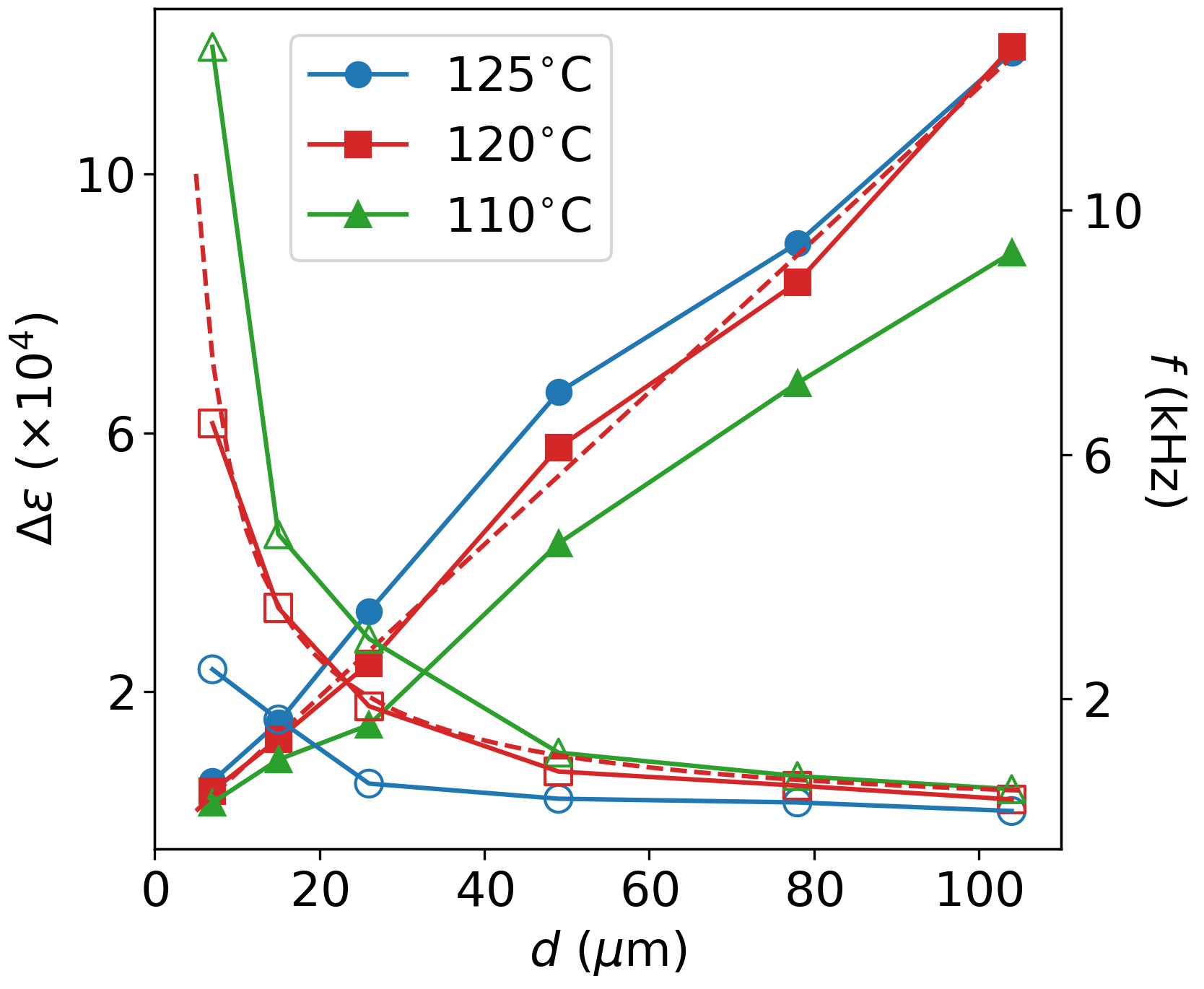}
\end{center}
\caption{\label{fig:fitpcg} Relaxation amplitudes (filled symbols) and frequencies (empty symbols) of the characteristic relaxation process at three temperatures of the N$_{\mathrm{F}}$ vs. cell thickness. The red dashed lines are fits to the experimental data at $120^{\circ}$C.}
\end{figure}

The model proposed in a recent preprint by Clark et al. \cite{clark_dielectric_2022} states that when a voltage is applied across a measuring cell with polymer aligning layers (high-capacity interfacial layers), the polarization vector, which is initially uniform and parallel to the plates, reorients, and with this reorientation the applied field is cancelled in the bulk sample. The N$_{\mathrm{F}}$ layer would then become, in an effective manner, conductive, enabling the charging of the interfacial capacitors. A so-called polarization-external capacitance  Goldstone (PCG) reorientation mode is then the result of the coupling of orientation and charge. The model predicts a relaxation process with amplitude $\Delta\varepsilon_\mathrm{PCG}\propto d$ and frequency $f_\mathrm{PCG}\propto 1/d$. Fig. \ref{fig:fitpcg} shows $\Delta\varepsilon$ and $f$ as a function of thickness obtained from fits to the experimental data. The dependence on $d$ mentioned above are well reproduced by the fitting results, although it should be mentioned that this was done considering a HN relaxation process and not a simple Debye relaxation process. It is also interesting to note that the distribution of relaxation times also depends on $d$, as evidenced by the change from 0.91 to 0.76 of the parameter $\alpha$, which defines the broadness of the process (see Eq. S1 and Table S1). In the description of the PCG model, it is pointed out that the disappearance in the N$_{\mathrm{F}}$ phase of the ionic-conductivity characteristic of the imaginary component with a 1/$\omega$ frequency dependence \cite{clark_dielectric_2022} can be used as an additional test. Fig. \ref{fig:measured_curves}.b shows that this is not the case in these measurements. However, is it important to recall at this point that the measurements presented in this paper were performed without alignment layers, i.e. with the material in contact with the gold electrodes.

Varying anchoring strength conditions, including the use of bare gold electrodes, have also been covered by Vaupotič et al. \cite{vaupotic_dielectric_2022}, who noted that the strength of the PCG mode should increase with increasing bias field, although the effect is not seen in the case of their measurements. Our results do not show this effect either. In terms of their CPM \cite{vaupotic_dielectric_2022}, the dielectric spectrum is described by two relaxation processes: a polarization reorientation mode (referred to as phason mode) and a polarization amplitude mode (amplitudon). At 0 DC bias field, the main relaxation process of the N$_{\mathrm{F}}$ phase would be dominated by the phason. However, according to their experimental findings, upon the application of a DC bias field the phason is suppressed and shifted towards low frequencies, while the amplitudon becomes visible and moves to higher frequencies \cite{vaupotic_dielectric_2022}. This appears to be compatible with our results, since the $8$ V$_{\mathrm{DC}}$ curve reveals a process of much lower amplitude and slightly higher frequency, while the low-frequency increase in $\varepsilon'$ could be attributed to the shifted phason (see Fig. \ref{fig:ACDC}). Finally, although a possible explanation for the decrease of $\varepsilon'$ at high frequencies is the suppression of a non-collective molecular mode, one should not rule out the possibility that it is the high-frequency collective mode predicted for the N$_{\mathrm{F}}$ phase by this same model, whose frequency is expected to be in the MHz regime \cite{vaupotic_dielectric_2022}. Returning to the thickness dependence, it is shown that the frequencies of the ferroelectric modes depend on the cell thickness as $\xi_K^2/d^2$, $\xi_b^2/d^2$ and $\xi_\gamma/d$, where $\xi_K$, $\xi_b$ and $\xi_\gamma$ are the director elastic, polarization elastic and flexoelectric characteristic lengths respectively. These terms, calculated either from assumed flexoelectric coefficient values ($\xi_\gamma/d$) or in base of their results ($\xi_b^2/d^2$), are considered negligible, and no dependence of the frequencies with $d$ is further described. However, our results clearly show such dependence, in line with the flexoelectric term $\xi_{\mathrm{\gamma}}/d$ (Fig. \ref{fig:fitpcg}). It is interesting to note that Barthakur et al. have recently reported a ``colossal flexoelectric effect" in RM734 above the nonpolar to polar nematic phase transition \cite{flexoelectric}. All of this indicates that the development of experimental approaches for an accurate determination of material parameters in the ferroelectric nematic phase, such as flexoelectric coefficients, elastic constants or polarization elasticity, is required for a deeper interpretation of the results. Finally, the previously mentioned dependence of the distribution of relaxation times of the dielectric mode ($\alpha$) on $d$ could perhaps be related to a deeper dependence of these material parameters.

\subsection{Analysis in terms of EP}
\begin{figure}[t!]
\begin{center}
\includegraphics[width=78 mm]{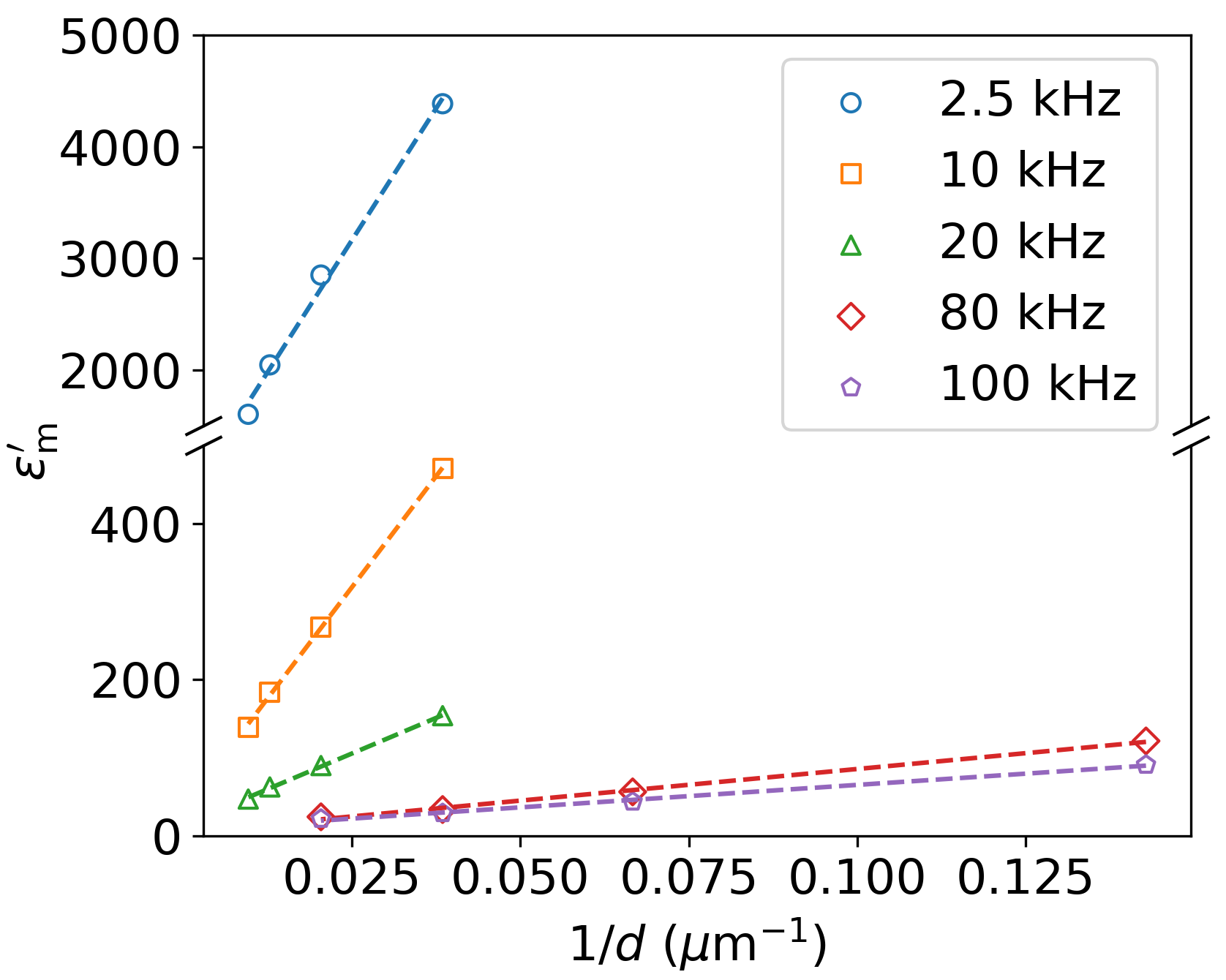}
\end{center}
\caption{\label{fig:eps_vs_d-1} Measured permittivity vs. inverse of electrode separation at different frequencies at $125^{\circ}$C. Only datapoints with common linear behavior are shown.}
\end{figure}

The fact that at low frequencies in the N phase, the real component of the permittivity increases, which evidences the presence of EP effects, and that at the transition to the polar phase, the ionic conductivity contribution does not disappear (Fig. \ref{fig:3dplot}), stimulates us to analyze our results in the N$_{\mathrm{F}}$ phase in terms of EP models. When an electric field is applied to the sample, free ions tend to move towards the electrode-sample interface giving rise to electrical double layers. The voltage drops rapidly in these regions, which leads to a large polarization and effective screening of the electric field in the bulk sample. This can result in extremely high values of the real and imaginary components of the dielectric permittivity at low frequencies. This is a well-known effect thoroughly investigated for a wide variety of systems \cite{european}, and it is also present in nematic liquid crystals as observed, for example, in materials such as 5CB in different conditions \cite{ shuichi_murakami_electrode_1997, murakami_dielectric_1996, 5cb_prakash}. Indeed, EP effects result in a similar linear dependence on $d$ of the measured $\varepsilon$ at ultralow frequencies for 5CB investigated with bare ITO electrodes \cite{murakami_dielectric_1996}. In the case of N$_{\mathrm{F}}$ materials, ion concentration was estimated to be up to several orders of magnitude larger than in conventional nematics \cite{basnet_soliton_2022} and, thus,  EP characteristic frequencies are expected to be higher than in the latter. Over the years, a great number of models have been developed to describe EP and here we will, as an example, analyze our data in terms of an ion diffusion model by Chassagne \emph{et al.} \cite{chassagne_compensating_2016}. Even though the authors intend to use this model for dielectric studies of colloidal suspensions, it can also be applied to the present case, since no assumption on the nature of the constituent particles is made. Therein, a method to compensate for EP based on using cells of different thicknesses is proposed as well. In this framework, the spectra are characterized by three (angular) frequencies: $\omega_{\mathrm{P}}$, below which ions fully accumulate at the electrodes; $\omega_{\text{0}}$, above which EP effects become negligible; and an intermediate frequency $\omega_{\text{b}}=\sqrt{\omega_{\text{P}}\omega_{\text{0}}}$. In the high frequency limit, $\omega>\omega_{\text{b}}$, the measurement is no longer affected by this phenomenon and it reflects the real material value. Below $\omega_{\mathrm{P}}$, the model predicts a linear dependence of the measured permittivity $\varepsilon'_\mathrm{m}$ (vs. the intrinsic sample permittivity $\varepsilon'_\mathrm{LC}$) with $d$, compatible with our experimental results (qualitatively this can be seen in Fig. \ref{fig:measured_curves}). Intuitively, this can be understood as follows: as the thickness of the measuring cell is increased, the ions can move further and, as a result, higher polarization is created and hence a higher permittivity is measured. In the intermediate frequency region, $\omega_{\mathrm{P}}<< \omega <<\omega_{\mathrm{0}}$, $\varepsilon'_\mathrm{m}$ depends on the thickness $d$ as:

\begin{equation}
\label{for:chassagintermediatemain}
    \varepsilon'_{\mathrm{m}}(\omega)=\frac{2\sigma_{\mathrm{LC}}^2}{\varepsilon_0^2 \varepsilon^* \kappa}\cdot\frac{1}{\omega^2}\cdot\frac{1}{d} + \varepsilon'_{\mathrm{LC}}(\omega),
\end{equation}

\noindent where $\sigma_{\mathrm{LC}}$ is the conductivity of the sample, $\varepsilon_0$ is the vacuum permittivity, $\varepsilon^*$ is a permittivity related to EP and $\kappa$ is the inverse of the Debye length. The characteristic frequencies $\omega_{\mathrm{P}}, \omega_{\mathrm{b}}$ and $\omega_{\mathrm{0}}$ are identified in Fig. S4 for the case of the 104 $\mu$m thick cell. It is evident that, as expected, such frequencies vary with thickness. From Eq. (\ref{for:chassagintermediatemain}) it is possible to obtain $\varepsilon'_{\mathrm{LC}}(\omega)$ at each frequency point directly from the intercept of the linear regression of the measured permittivity for different thicknesses. Fig. \ref{fig:eps_vs_d-1} shows regression examples at several frequencies. It is important to note that this correction method is only justified in the intermediate frequency region, which is different for each thickness. Therefore, at a given $\omega$, only the data for some of the thicknesses can be employed (see Fig. \ref{fig:eps_vs_d-1}).

\begin{figure}[t!]
\begin{center}
\includegraphics[width=70 mm]{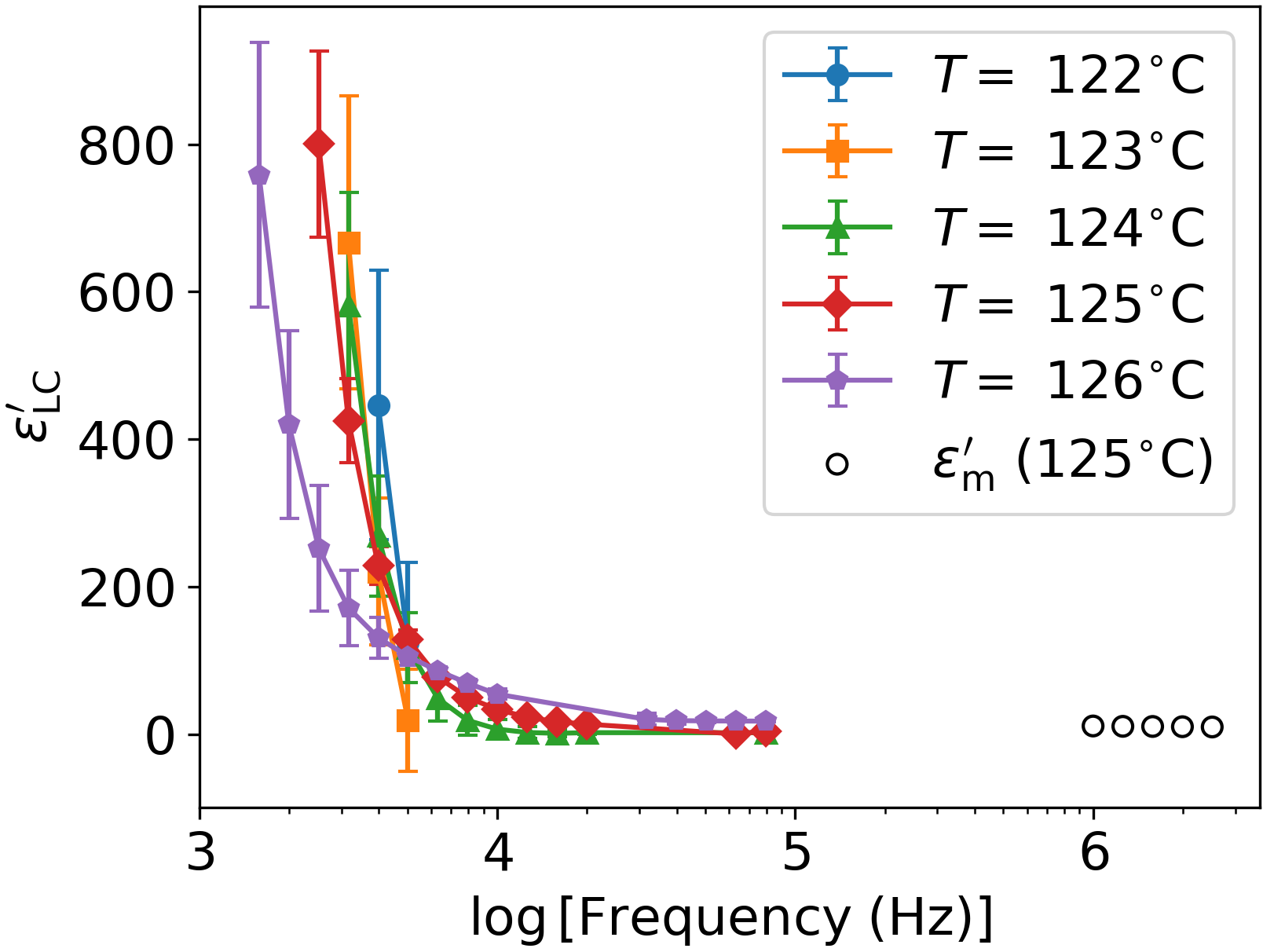}
\end{center}
\caption{\label{fig:eps_corrected} $\varepsilon'_\mathrm{LC}$ vs. frequency obtained as EP corrected data (solid symbols). Directly measured data $\varepsilon'_{\mathrm{m}}=\varepsilon'_{\mathrm{LC}}$ in the EP free region (empty symbols) at $125^{\circ}$C (these EP free values at high frequency do not appreciably vary at the rest of the temperatures shown).}
\end{figure}

In this way, $\varepsilon'_{\mathrm{LC}}$ for EP corrected spectra can be deduced at frequencies within the validity range. Results at several temperatures are shown in Fig. \ref{fig:eps_corrected}. This suggests the existence of a relaxation process below 1 kHz with an amplitude larger than $10^3$. The value obtained in this calculation at higher frequencies ($\sim10$) is in agreement with the EP-free values of Fig. \ref{fig:measured_curves} at high frequencies. Due to the comparatively large value of the signal vs. $\varepsilon_\mathrm{LC}'$, these results should be taken as estimates. A comparison of EP corrected data vs. measured permittivity can be found in Fig. S5. It should be noted that strict reliability criteria were selected for the EP analysis. Only frequencies for which at least four data points were available for the linear regressions were analyzed (see Fig. \ref{fig:eps_vs_d-1}), and the latter were only further considered if $R^2>0.99$ was met. Under these criteria, EP correction only resulted in physically relevant values (i.e. $\varepsilon_\mathrm{LC}'> 0$) in a limited temperature range above $\sim 120^\circ$C.

Finally, it should be noted that, according to the EP model, the slopes obtained from the linear regressions in the intermediate frequency region should have a $1/f^{\:n=2}$ dependence (see formula \ref{for:chassagintermediatemain}). However, in our case $n$ varies between $1.4$ and $1.7$ on cooling, calculated as described in Fig. S6. Some authors have suggested that this deviation is to be expected under certain conditions when indeed $n=3/2$ should be observed \cite{chassagne_compensating_2016,cirkel,hollingsworth}. In any case, the broadening of the curves is related to the above-mentioned behavior of the $\alpha$ parameters obtained in the HN fits and calls for taking the results with care. 

\section{Conclusions}
\label{sec:discussion}
The systematic investigation of the dielectric behavior of RM734 for different measuring conditions presented in this work (i.e. thicknesses of the measuring parallel plate capacitor) experimentally evidences that an adequate interpretation of dielectric spectroscopy measurements in ferroelectric nematic liquid crystals requires the development of dedicated models. The permittivity values obtained here are in agreement with those reported for several N$_\mathrm{F}$ materials (e.g. \cite{li_development_2021, manabe_ferroelectric_2021, yadav_polar_2022, vaupotic_dielectric_2022}), and thus similar thickness effects can be expected as a general rule for these complex fluids.

The results have been analyzed in the light of three different models: the PCG \cite{clark_dielectric_2022}, the CPM \cite{vaupotic_dielectric_2022} and the EP \cite{chassagne_compensating_2016} models. The three account for the dependence of the measured permittivity on the sample thickness; however, the underlying mechanisms greatly differ. The PCG model takes into account the large polarization values of the ferroelectric nematic materials and the cancellation of external fields, such as the AC measuring field, resulting from the small reorientations of such large polarization. In the EP model, large values of permittivity are expected at low frequencies arising from the ionic conductivity of the studied sample. In the CPM model, the thickness dependence is related to the flexoelectricity, nematic elasticity and large polarization of the materials. Determining the relevant material parameters that are currently absent could help to interpret the results in terms of this model.

It would be reasonable to assume that in N$_\mathrm{F}$ systems, more than one of these mechanisms can be involved. Additionally, it should be taken into account that polarization structures in confined ferroelectric nematic materials can be intricate, with the formation of different domains divided by walls and defects. The latter can additionally carry electrical charge. Understanding the origin of the large permittivities in the N$_\mathrm{F}$ phase is key both from fundamental and application-focused point of views. Further experimental research in this direction should focus, on the one hand, on systematic research of materials other than RM734 in the hope of obtaining more distinctive results. On the other hand, the employment of measurement methods of $\varepsilon(\omega)$ free of EP, such as open-ended four-electrode measurements, despite being more difficult to implement in the laboratory, could shed some light on the mechanisms at play.

\section*{Acknowledgements}
A.E. and J.M.-P. acknowledge funding from the Basque Goverment Project IT1458-22. A.E. thanks the Department of Education of the Basque Government for a predoctoral fellowship (grant no. PRE\_2022\_1\_0104). R.J.M. thanks UKRI for funding via a Future Leaders Fellowship (grant no. MR/W006391/1). N.S. and A.M. acknowledge financial support from the Slovenian Research Agency (research core Funding No. P1-0192).

\appendix

\bibliographystyle{elsarticle-num} 
\bibliography{REFS}

\begin{thebibliography}{10}
\expandafter\ifx\csname url\endcsname\relax
  \def\url#1{\texttt{#1}}\fi
\expandafter\ifx\csname urlprefix\endcsname\relax\def\urlprefix{URL }\fi
\expandafter\ifx\csname href\endcsname\relax
  \def\href#1#2{#2} \def\path#1{#1}\fi

\bibitem{debye}
P.~Debye, {Einige Resultate einer kinetischen Theorie der Isolatoren},
  Physikalische Zeitschrift 13 (1912) 97--100.

\bibitem{born}
M.~Born, {Über anisotrope Flüssigkeiten: Versuch einer Theorie der flüssigen
  Kristalle und des elektrischen Kerr-Effekts in Flüssigkeiten},
  Sitzungsberichte der Preussischen Akademie der Wissenschaften 30 (1916)
  614--650.

\bibitem{mandle_nematic_2017}
R.~J. Mandle, S.~J. Cowling, J.~W. Goodby, A nematic to nematic transformation
  exhibited by a rod-like liquid crystal, Physical Chemistry Chemical Physics
  19~(18) (2017) 11429--11435.
\newblock \href {https://doi.org/10.1039/C7CP00456G}
  {\path{doi:10.1039/C7CP00456G}}.

\bibitem{nishikawa18}
H.~Nishikawa, K.~Shiroshita, H.~Higuchi, Y.~Okumura, Y.~Haseba, S.-i. Yamamoto,
  K.~Sago, H.~Kikuchi, A fluid liquid-crystal material with highly polar order,
  Advanced Materials 29~(43) (2017) 1702354.
\newblock \href {https://doi.org/10.1002/adma.201702354}
  {\path{doi:10.1002/adma.201702354}}.

\bibitem{chen_first-principles_2020}
X.~Chen, E.~Korblova, D.~Dong, X.~Wei, R.~Shao, L.~Radzihovsky, M.~A. Glaser,
  J.~E. Maclennan, D.~Bedrov, D.~M. Walba, N.~A. Clark, First-principles
  experimental demonstration of ferroelectricity in a thermotropic nematic
  liquid crystal: {Polar} domains and striking electro-optics, Proceedings of
  the National Academy of Sciences 117~(25) (2020) 14021--14031.
\newblock \href {https://doi.org/10.1073/pnas.2002290117}
  {\path{doi:10.1073/pnas.2002290117}}.

\bibitem{chen_ideal_2022}
X.~Chen, Z.~Zhu, M.~J. Magrini, E.~Korblova, C.~S. Park, M.~A. Glaser, J.~E.
  Maclennan, D.~M. Walba, N.~A. Clark, Ideal mixing of paraelectric and
  ferroelectric nematic phases in liquid crystals of distinct molecular
  species, Liquid Crystals 49~(11) (2022) 1531--1544.
\newblock \href {https://doi.org/10.1080/02678292.2022.2058101}
  {\path{doi:10.1080/02678292.2022.2058101}}.

\bibitem{brown_multiple_2021}
S.~Brown, E.~Cruickshank, J.~M.~D. Storey, C.~T. Imrie, D.~Pociecha,
  M.~Majewska, A.~Makal, E.~Gorecka, Multiple {Polar} and {Non}‐polar
  {Nematic} {Phases}, ChemPhysChem 22~(24) (2021) 2506--2510.
\newblock \href {https://doi.org/10.1002/cphc.202100644}
  {\path{doi:10.1002/cphc.202100644}}.

\bibitem{folcia_ferroelectric_2022}
C.~L. Folcia, J.~Ortega, R.~Vidal, T.~Sierra, J.~Etxebarria, The ferroelectric
  nematic phase: an optimum liquid crystal candidate for nonlinear optics,
  Liquid Crystals 49~(6) (2022) 899--906.
\newblock \href {https://doi.org/10.1080/02678292.2022.2056927}
  {\path{doi:10.1080/02678292.2022.2056927}}.

\bibitem{li_how_2021}
J.~Li, R.~Xia, H.~Xu, J.~Yang, X.~Zhang, J.~Kougo, H.~Lei, S.~Dai, H.~Huang,
  G.~Zhang, F.~Cen, Y.~Jiang, S.~Aya, M.~Huang, How {Far} {Can} {We} {Push} the
  {Rigid} {Oligomers}/{Polymers} toward {Ferroelectric} {Nematic} {Liquid}
  {Crystals}?, Journal of the American Chemical Society 143~(42) (2021)
  17857--17861.
\newblock \href {https://doi.org/10.1021/jacs.1c09594}
  {\path{doi:10.1021/jacs.1c09594}}.

\bibitem{li_development_2021}
J.~Li, H.~Nishikawa, J.~Kougo, J.~Zhou, S.~Dai, W.~Tang, X.~Zhao, Y.~Hisai,
  M.~Huang, S.~Aya, Development of ferroelectric nematic fluids with
  giant-$\varepsilon$ dielectricity and nonlinear optical properties, Science
  Advances 7~(17) (2021) eabf5047.
\newblock \href {https://doi.org/10.1126/sciadv.abf5047}
  {\path{doi:10.1126/sciadv.abf5047}}.

\bibitem{manabe_ferroelectric_2021}
A.~Manabe, M.~Bremer, M.~Kraska, Ferroelectric nematic phase at and below room
  temperature, Liquid Crystals 48~(8) (2021) 1079--1086.
\newblock \href {https://doi.org/10.1080/02678292.2021.1921867}
  {\path{doi:10.1080/02678292.2021.1921867}}.

\bibitem{yadav_polar_2022}
N.~Yadav, Y.~P. Panarin, J.~K. Vij, W.~Jiang, G.~H. Mehl, Two mechanisms for
  the formation of the ferronematic phase studied by dielectric spectroscopy,
  Journal of Molecular Liquids 378 (2023) 121570.
\newblock \href {https://doi.org/10.1016/j.molliq.2023.121570}
  {\path{doi:10.1016/j.molliq.2023.121570}}.

\bibitem{nishikawa_new_2021}
H.~Nishikawa, F.~Araoka, A {New} {Class} of {Chiral} {Nematic} {Phase} with
  {Helical} {Polar} {Order}, Advanced Materials 33~(35) (2021) 2101305.
\newblock \href {https://doi.org/10.1002/adma.202101305}
  {\path{doi:10.1002/adma.202101305}}.

\bibitem{nishikawa_anisotropic_2022}
H.~Nishikawa, K.~Sano, F.~Araoka, Anisotropic fluid with phototunable
  dielectric permittivity, Nature Communications 13~(1) (2022) 1142.
\newblock \href {https://doi.org/10.1038/s41467-022-28763-1}
  {\path{doi:10.1038/s41467-022-28763-1}}.

\bibitem{clark_dielectric_2022}
N.~A. Clark, X.~Chen, J.~E. Maclennan, M.~A. Glaser, Dielectric spectroscopy of
  ferroelectric nematic liquid crystals: {Measuring} the capacitance of
  insulating interfacial layers, arXiv:2208.09784 [cond-mat] (Aug. 2022).
\newblock \href {https://doi.org/10.48550/arXiv.2208.09784}
  {\path{doi:10.48550/arXiv.2208.09784}}.

\bibitem{vaupotic_dielectric_2022}
N.~Vaupotič, D.~Pociecha, P.~Rybak, J.~Matraszek, M.~Čepič, J.~M. Wolska,
  E.~Gorecka, Dielectric response of a ferroelectric nematic liquid crystalline
  phase in thin cells, Liquid Crystals (2023) 1--12\href
  {https://doi.org/10.1080/02678292.2023.2180099}
  {\path{doi:10.1080/02678292.2023.2180099}}.

\bibitem{zhao_spontaneous_2021}
X.~Zhao, J.~Zhou, J.~Li, J.~Kougo, Z.~Wan, M.~Huang, S.~Aya, Spontaneous
  helielectric nematic liquid crystals: {Electric} analog to helimagnets,
  Proceedings of the National Academy of Sciences 118~(42) (2021) e2111101118.
\newblock \href {https://doi.org/10.1073/pnas.2111101118}
  {\path{doi:10.1073/pnas.2111101118}}.

\bibitem{chiral_tunable}
C.~Feng, R.~Saha, E.~Korblova, D.~Walba, S.~N. Sprunt, A.~Jákli, Electrically
  tunable reflection color of chiral ferroelectric nematic liquid crystals,
  Advanced Optical Materials 9~(22) (2021) 2101230.
\newblock \href {https://doi.org/10.1002/adom.202101230}
  {\path{doi:10.1002/adom.202101230}}.

\bibitem{ortega_ferroelectric_2022}
J.~Ortega, C.~L. Folcia, J.~Etxebarria, T.~Sierra, Ferroelectric chiral nematic
  liquid crystals: new photonic materials with multiple bandgaps controllable
  by low electric fields, Liquid Crystals 49~(15) (2022) 2128--2136.
\newblock \href {https://doi.org/10.1080/02678292.2022.2104949}
  {\path{doi:10.1080/02678292.2022.2104949}}.

\bibitem{POCIECHA2022119532}
D.~Pociecha, R.~Walker, E.~Cruickshank, J.~Szydlowska, P.~Rybak, A.~Makal,
  J.~Matraszek, J.~M. Wolska, J.~M. Storey, C.~T. Imrie, E.~Gorecka,
  Intrinsically chiral ferronematic liquid crystals: An inversion of the
  helical twist sense at the chiral nematic – chiral ferronematic phase
  transition, Journal of Molecular Liquids 361 (2022) 119532.
\newblock \href {https://doi.org/10.1016/j.molliq.2022.119532}
  {\path{doi:10.1016/j.molliq.2022.119532}}.

\bibitem{chen_observation_2022}
X.~Chen, V.~Martinez, P.~Nacke, E.~Korblova, A.~Manabe, M.~Klasen-Memmer,
  G.~Freychet, M.~Zhernenkov, M.~A. Glaser, L.~Radzihovsky, J.~E. Maclennan,
  D.~M. Walba, M.~Bremer, F.~Giesselmann, N.~A. Clark, Observation of a
  uniaxial ferroelectric smectic {A} phase, Proceedings of the National Academy
  of Sciences 119~(47) (2022) e2210062119.
\newblock \href {https://doi.org/10.1073/pnas.2210062119}
  {\path{doi:10.1073/pnas.2210062119}}.

\bibitem{kikuchi_fluid_2022}
H.~Kikuchi, H.~Matsukizono, K.~Iwamatsu, S.~Endo, S.~Anan, Y.~Okumura, Fluid
  {Layered} {Ferroelectrics} with {Global} {C}$_{\mathrm{\infty v}}$
  {Symmetry}, Advanced Science 9~(26) (2022) 2202048.
\newblock \href {https://doi.org/10.1002/advs.202202048}
  {\path{doi:10.1002/advs.202202048}}.

\bibitem{shen_effective_2011}
Y.~Shen, T.~Gong, R.~Shao, E.~Korblova, J.~E. Maclennan, D.~M. Walba, N.~A.
  Clark, Effective conductivity due to continuous polarization reorientation in
  fluid ferroelectrics, Physical Review E 84~(2) (2011) 020701.
\newblock \href {https://doi.org/10.1103/PhysRevE.84.020701}
  {\path{doi:10.1103/PhysRevE.84.020701}}.

\bibitem{kremer_broadband_2003}
F.~Kremer, A.~Schönhals (Eds.), Broadband {Dielectric} {Spectroscopy},
  Springer Berlin Heidelberg, Berlin, Heidelberg, 2003.
\newblock \href {https://doi.org/10.1007/978-3-642-56120-7}
  {\path{doi:10.1007/978-3-642-56120-7}}.

\bibitem{chassagne_compensating_2016}
C.~Chassagne, E.~Dubois, M.~L. Jiménez, J.~P.~M. van~der Ploeg, J.~van
  Turnhout, Compensating for {Electrode} {Polarization} in {Dielectric}
  {Spectroscopy} {Studies} of {Colloidal} {Suspensions}: {Theoretical}
  {Assessment} of {Existing} {Methods}, Frontiers in Chemistry 4 (2016).

\bibitem{basnet_soliton_2022}
B.~Basnet, M.~Rajabi, H.~Wang, P.~Kumari, K.~Thapa, S.~Paul, M.~O.
  Lavrentovich, O.~D. Lavrentovich, Soliton walls paired by polar surface
  interactions in a ferroelectric nematic liquid crystal, Nature Communications
  13~(1) (2022) 3932.
\newblock \href {https://doi.org/10.1038/s41467-022-31593-w}
  {\path{doi:10.1038/s41467-022-31593-w}}.

\bibitem{patterning}
N.~Sebasti{\'{a}}n, M.~Lov{\v{s}}in, B.~Berteloot, N.~Osterman, A.~Petelin,
  R.~J. Mandle, S.~Aya, M.~Huang, I.~Dreven{\v{s}}ek-Olenik, K.~Neyts,
  A.~Mertelj, Polarization patterning in ferroelectric nematic liquids via
  flexoelectric coupling, Nature Communications 14~(1) (2023).
\newblock \href {https://doi.org/10.1038/s41467-023-38749-2}
  {\path{doi:10.1038/s41467-023-38749-2}}.

\bibitem{mandle_rational_2017}
R.~J. Mandle, S.~J. Cowling, J.~W. Goodby, Rational {Design} of {Rod}-{Like}
  {Liquid} {Crystals} {Exhibiting} {Two} {Nematic} {Phases}, Chemistry - A
  European Journal 23~(58) (2017) 14554--14562.
\newblock \href {https://doi.org/10.1002/chem.201702742}
  {\path{doi:10.1002/chem.201702742}}.

\bibitem{mertelj_splay_2018}
A.~Mertelj, L.~Cmok, N.~Sebastián, R.~J. Mandle, R.~R. Parker, A.~C. Whitwood,
  J.~W. Goodby, M.~Čopič, Splay {Nematic} {Phase}, Physical Review X 8~(4)
  (2018) 041025.
\newblock \href {https://doi.org/10.1103/PhysRevX.8.041025}
  {\path{doi:10.1103/PhysRevX.8.041025}}.

\bibitem{cut_off}
P.~Perkowski, The parasitic effects in high-frequency dielectric spectroscopy
  of liquid crystals – the review, Liquid Crystals 48~(6) (2021) 767--793.
\newblock \href {https://doi.org/10.1080/02678292.2020.1852619}
  {\path{doi:10.1080/02678292.2020.1852619}}.

\bibitem{shuichi_murakami_electrode_1997}
S.~M. Shuichi~Murakami, H.~N. Hiroyoshi~Naito, Electrode and {Interface}
  {Polarizations} in {Nematic} {Liquid} {Crystal} {Cells}, Japanese Journal of
  Applied Physics 36~(4R) (1997) 2222.
\newblock \href {https://doi.org/10.1143/JJAP.36.2222}
  {\path{doi:10.1143/JJAP.36.2222}}.

\bibitem{5cb}
K.~Kondratenko, Y.~Boussoualem, S.~Longuemart, A.~Daoudi, Ionic transport in
  nematic liquid crystals and alignment layer effects on electrode
  polarization, The Journal of Chemical Physics 149~(13) (2018) 134902.
\newblock \href {https://doi.org/10.1063/1.5045268}
  {\path{doi:10.1063/1.5045268}}.

\bibitem{PhysRevLett.124.037801}
N.~Sebasti\'an, L.~Cmok, R.~J. Mandle, M.~R. de~la Fuente, I.~Dreven\ifmmode
  \check{s}\else~\v{s}\fi{}ek Olenik, M.~\ifmmode \check{C}\else
  \v{C}\fi{}opi\ifmmode~\check{c}\else \v{c}\fi{}, A.~Mertelj,
  Ferroelectric-ferroelastic phase transition in a nematic liquid crystal,
  Physical Review Letters 124 (2020) 037801.
\newblock \href {https://doi.org/10.1103/PhysRevLett.124.037801}
  {\path{doi:10.1103/PhysRevLett.124.037801}}.

\bibitem{mandle_molecular_2021}
R.~J. Mandle, N.~Sebastián, J.~Martinez-Perdiguero, A.~Mertelj, On the
  molecular origins of the ferroelectric splay nematic phase, Nature
  Communications 12~(1) (2021) 4962.
\newblock \href {https://doi.org/10.1038/s41467-021-25231-0}
  {\path{doi:10.1038/s41467-021-25231-0}}.

\bibitem{murakami_dielectric_1996}
S.~Murakami, H.~Iga, H.~Naito, Dielectric properties of nematic liquid crystals
  in the ultralow frequency regime, Journal of Applied Physics 80~(11) (1996)
  6396--6400.
\newblock \href {https://doi.org/10.1063/1.363658}
  {\path{doi:10.1063/1.363658}}.

\bibitem{5cb_prakash}
A.~Kumar, D.~Varshney, J.~Prakash, Role of ionic contribution in dielectric
  behaviour of a nematic liquid crystal with variable cell thickness, Journal
  of Molecular Liquids 303 (2020) 112520.
\newblock \href {https://doi.org/10.1016/j.molliq.2020.112520}
  {\path{doi:10.1016/j.molliq.2020.112520}}.

\bibitem{Saha_Nepal}
R.~Saha, P.~Nepal, C.~Feng, M.~S. Hossain, M.~Fukuto, R.~Li, J.~T. Gleeson,
  S.~Sprunt, R.~J. Twieg, A.~Jákli, Multiple ferroelectric nematic phases of a
  highly polar liquid crystal compound, Liquid Crystals 49~(13) (2022)
  1784--1796.
\newblock \href {https://doi.org/10.1080/02678292.2022.2069297}
  {\path{doi:10.1080/02678292.2022.2069297}}.

\bibitem{caimi}
F.~Caimi, G.~Nava, R.~Barboza, N.~A. Clark, E.~Korblova, D.~M. Walba,
  T.~Bellini, L.~Lucchetti, Surface alignment of ferroelectric nematic liquid
  crystals, Soft Matter 17 (2021) 8130--8139.
\newblock \href {https://doi.org/10.1039/D1SM00734C}
  {\path{doi:10.1039/D1SM00734C}}.

\bibitem{flexoelectric}
A.~Barthakur, J.~Karcz, P.~Kula, S.~Dhara, Critical splay fluctuations and
  colossal flexoelectric effect above the nonpolar to polar nematic phase
  transition, Phys. Rev. Mater. 7 (2023) 035603.
\newblock \href {https://doi.org/10.1103/PhysRevMaterials.7.035603}
  {\path{doi:10.1103/PhysRevMaterials.7.035603}}.

\bibitem{european}
S.~Emmert, M.~Wolf, R.~Gulich, S.~Krohns, S.~Kastner, P.~Lunkenheimer,
  A.~Loidl, Electrode polarization effects in broadband dielectric
  spectroscopy, The European Physical Journal B 83~(2) (2011) 157--165.
\newblock \href {https://doi.org/10.1140/epjb/e2011-20439-8}
  {\path{doi:10.1140/epjb/e2011-20439-8}}.

\bibitem{cirkel}
P.~Cirkel, J.~van~der Ploeg, G.~Koper, Electrode effects in dielectric
  spectroscopy of colloidal suspensions, Physica A: Statistical Mechanics and
  its Applications 235~(1-2) (1997) 269--278.
\newblock \href {https://doi.org/10.1016/S0378-4371(96)00347-0}
  {\path{doi:10.1016/S0378-4371(96)00347-0}}.

\bibitem{hollingsworth}
A.~D. Hollingsworth, Remarks on the determination of low-frequency measurements
  of the dielectric response of colloidal suspensions, Current Opinion in
  Colloid \& Interface Science 18~(2) (2013) 157--159.
\newblock \href {https://doi.org/10.1016/j.cocis.2013.01.002}
  {\path{doi:10.1016/j.cocis.2013.01.002}}.

\end{thebibliography}

\end{document}


\begin{frontmatter}



\title{SUPPLEMENTARY MATERIAL FOR\\Dielectric spectroscopy of a ferroelectric nematic liquid crystal and the effect of the sample thickness}


\author[inst1]{Aitor Erkoreka}

\affiliation[inst1]{organization={Department of Physics, Faculty of Science and Technology, University of the Basque Country UPV/EHU},
            city={Bilbao},
            country={Spain}}
\author[inst1]{Josu Martinez-Perdiguero}

\author[inst2,inst3] {Richard J. Mandle}

\affiliation[inst2]{organization={School of Physics and Astronomy, University of Leeds},
            city={Leeds},
            country={UK}}

\affiliation[inst3]{organization={School of Chemistry, University of Leeds},
            city={Leeds},
            country={UK}}
            
\author[inst4]{Alenka Mertelj}
\author[inst4]{Nerea Sebastián}

\affiliation[inst4]{organization={Jožef Stefan Institute},
            city={Ljubljana},
            country={Slovenia}}

\end{frontmatter}

\section*{Havriliak-Negami equation}
\label{sec:HN}

The Havriliak-Negami equation can be written as

\begin{equation}
    \varepsilon (\omega) = \sum_{k} \frac{\Delta \varepsilon_k}{\left[1+(i \omega \tau_{k})^{\alpha_k} \right]^{\beta_k}} + \varepsilon_{\infty} - i\frac{\sigma_{\mathrm{dc}}}{\omega \varepsilon_0},\label{HN_eq} \tag{S1}
\end{equation}

where $\Delta \varepsilon_k$, $\tau_k$, $\alpha_k$ and $\beta_k$ are respectively the dielectric strength, relaxation time and broadness exponents of mode $k$, $\varepsilon_{\infty}$ is the high frequency permittivity and $\sigma_{\mathrm{dc}}$ is the dc conductivity.

\renewcommand{\thefigure}{S1}
\begin{figure}[htbp]
\begin{center}
\includegraphics[width=80mm]{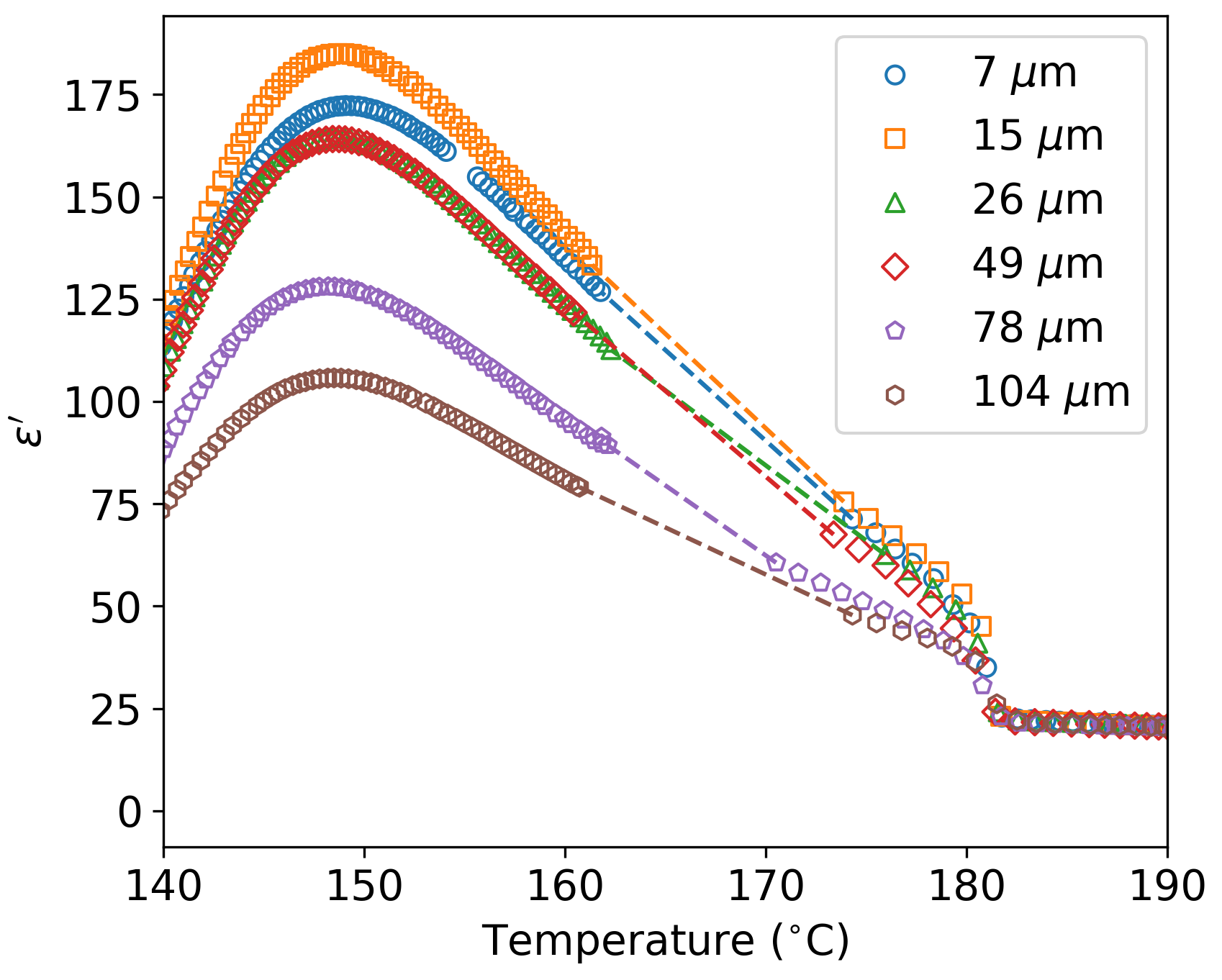}
\end{center}
\caption{\label{fig:tempDepiso} Temperature dependence of the measured dielectric permittivity at $100$ kHz for cells with different thicknesses showing the convergence in the isotropic phase. Dashed lines correspond just to visual guidelines.}
\end{figure}

\renewcommand{\thefigure}{S2}
\begin{figure}[htbp]
\begin{center}
\includegraphics[width=80mm]{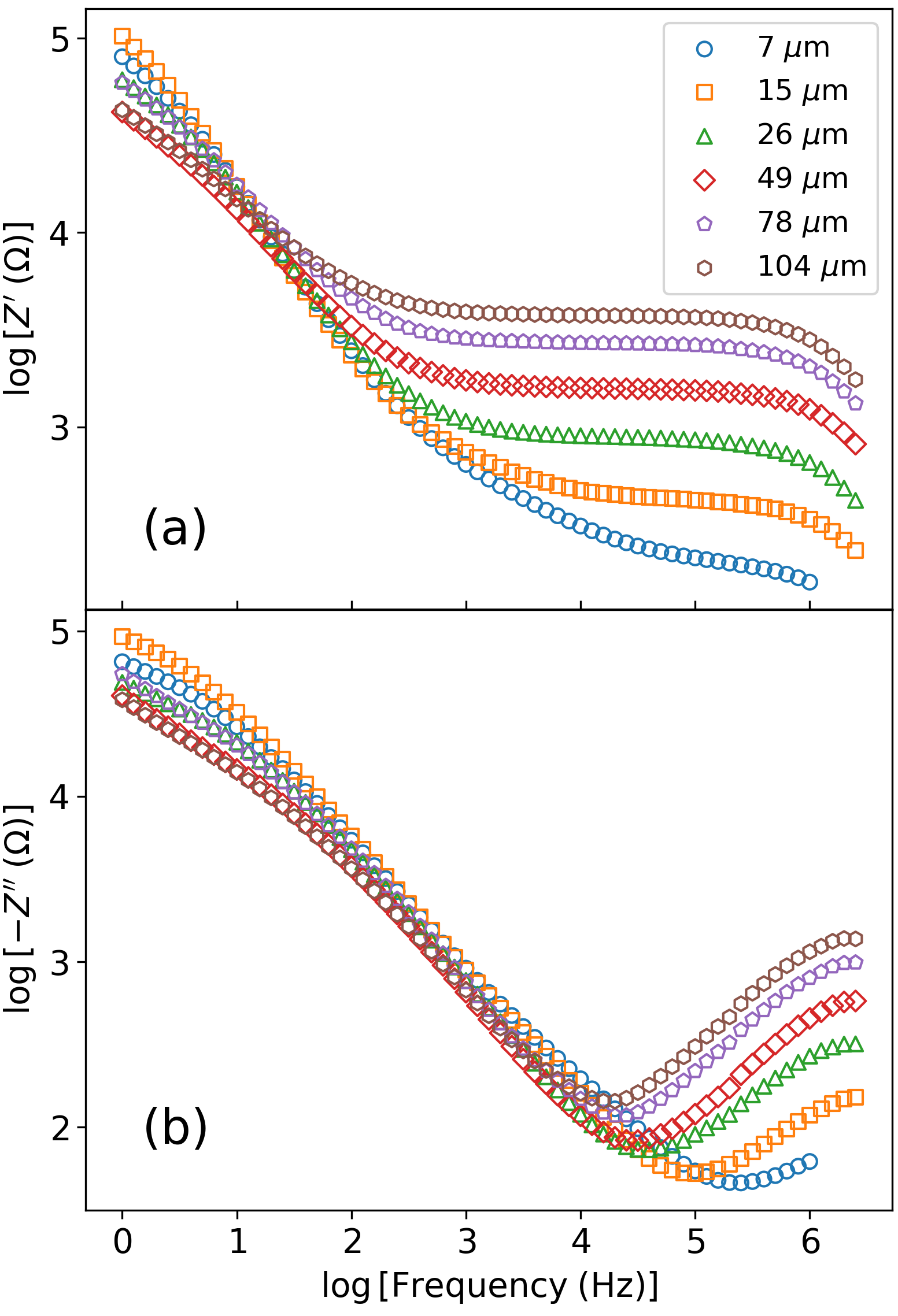}
\end{center}
\caption{\label{fig:impedance} Real (a) and imaginary (b) components of the complex impedance as a function of frequency for different electrode separations at $125^{\circ}$C.}
\end{figure}

\clearpage

\renewcommand{\thetable}{S1}
\begin{table}[H]
    \centering
    \begin{tabular}{|c|c|c|c|c|}
        \hline
        \multicolumn{1}{|c|}{} & \multicolumn{2}{c|}{N} & \multicolumn{2}{c|}{N$_\mathrm{F}$}\\
        \hline
        $d$ ($\mu$m) & $\alpha$ &  $\beta$ & $\alpha$ &  $\beta$ \\
        \hline
        7  & 0.97 - 0.87 & 1.00 & 0.76 - 0.82 & 1.15 \\
        \hline
        15 & 0.95 & 1.00 & 0.85 & 1.16 \\
        \hline
        26 & 0.95 & 1.00 & 0.85 & 1.17 \\
        \hline
        49 & 0.95 & 1.00 & 0.88 & 1.15 \\
        \hline
        78 & 0.87 - 0.93 & 1.16 - 0.98 & 0.91 & 1.15 \\
        \hline
        104 & 0.82 - 0.94 & 1.20 - 0.96 & 0.91 & 1.15 \\
        \hline
    \end{tabular}
    \caption{$\alpha$ and $\beta$ parameters obtained in the HN fittings with different thicknesses. The parameters were let free (with $\alpha<1$ and $\alpha\cdot\beta<1$) in the fitting procedure so that there are (small) fluctuations around the shown values. When a clear tendency was observed the intervals are specified (the values go from high to low temperature in the phase). $\Delta\varepsilon$ and the frequency of the relaxations are shown in Figure 6 of the main text.}
    \label{tab:example}
\end{table}

\renewcommand{\thefigure}{S3}
\begin{figure}[htbp]
\begin{center}
\includegraphics[width=130mm]{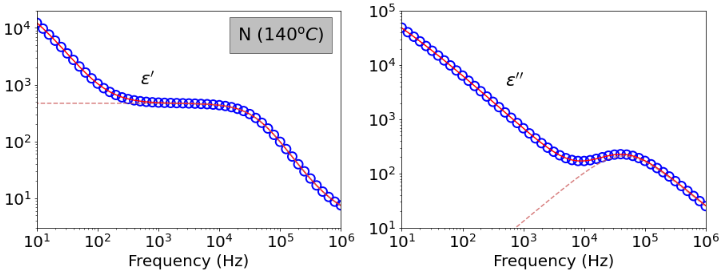}
\includegraphics[width=130mm]{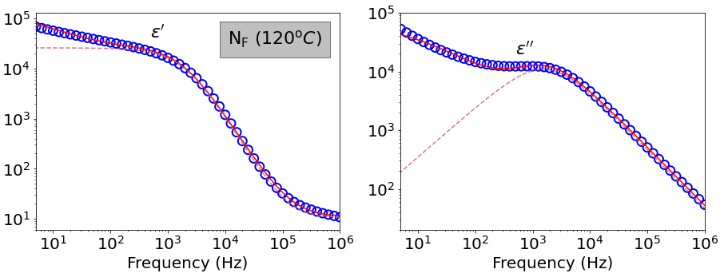}
\end{center}
\caption{\label{fig:fits} Examples of the perfomed fits in using equation S1. The data shown corresponds to the 26 $\mu$m thick cell. The full line corresponds to the full fit including the low-frequency contributions taken into account with a very low-frequency relaxation ($f<1$ Hz) and dc conductivity term. Dashed line corresponds to the contribution of the main mode (see Figure 6 in the main text). Top graphs: Real and imaginary parts of the permittivity in the N phase at $140^{\circ}$C. Bottom graphs: Real and imaginary parts of the permittivity in the N$_\mathrm{F}$ phase at $120^{\circ}$C.}
\end{figure}

\renewcommand{\thefigure}{S4}
\begin{figure}[htbp]
\begin{center}
\includegraphics[width=80mm]{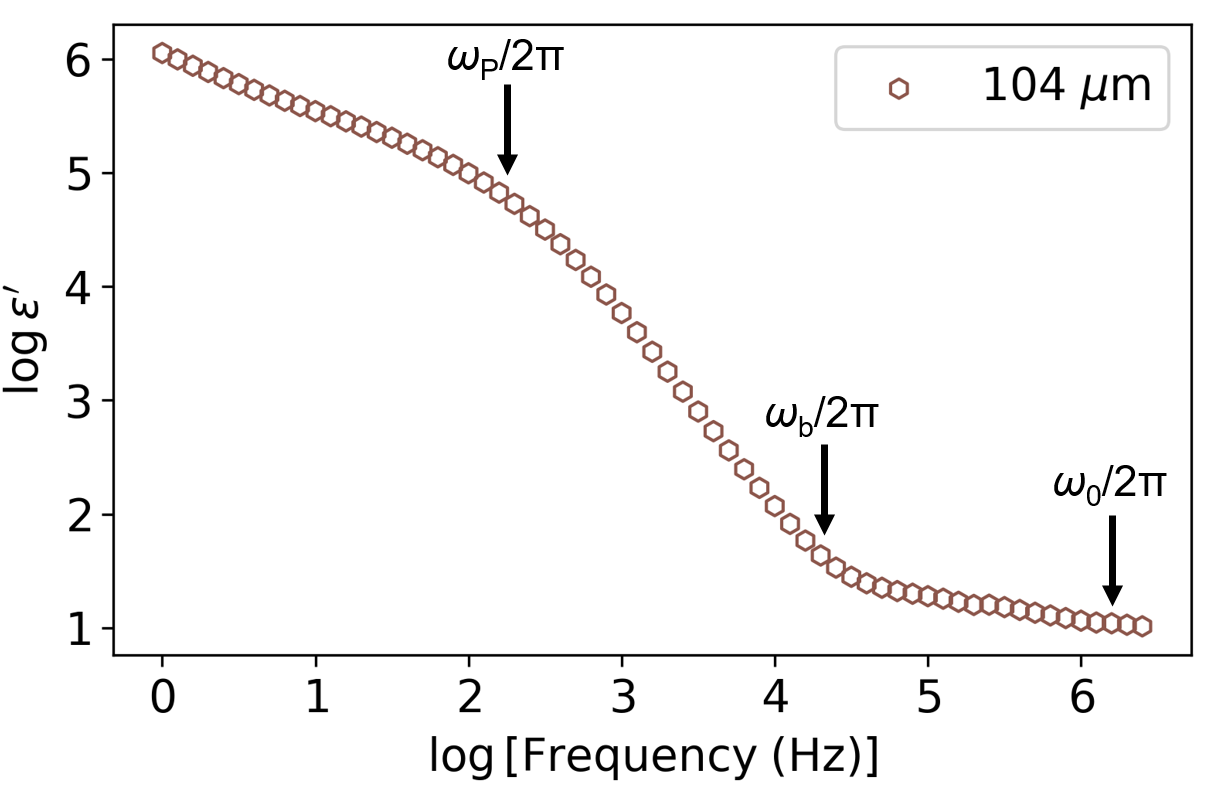}
\end{center}
\caption{\label{fig:identify_frequencies} Dielectric spectrum at $125^{\circ}$C for $d=104$ $\mu$m with the characteristic frequencies of the EP model: $\omega_{\mathrm{P}}/2\pi$, below which ions fully accumulate at the electrodes, $\omega_{\text{0}}/2\pi$, above which EP effects become negligible, and the intermediate frequency $\omega_{\text{b}}/2\pi=\sqrt{\omega_{\text{P}}\omega_{\text{0}}}/2\pi$.}
\end{figure}

\renewcommand{\thefigure}{S5}
\begin{figure}[htbp]
\begin{center}
\includegraphics[width=80mm]{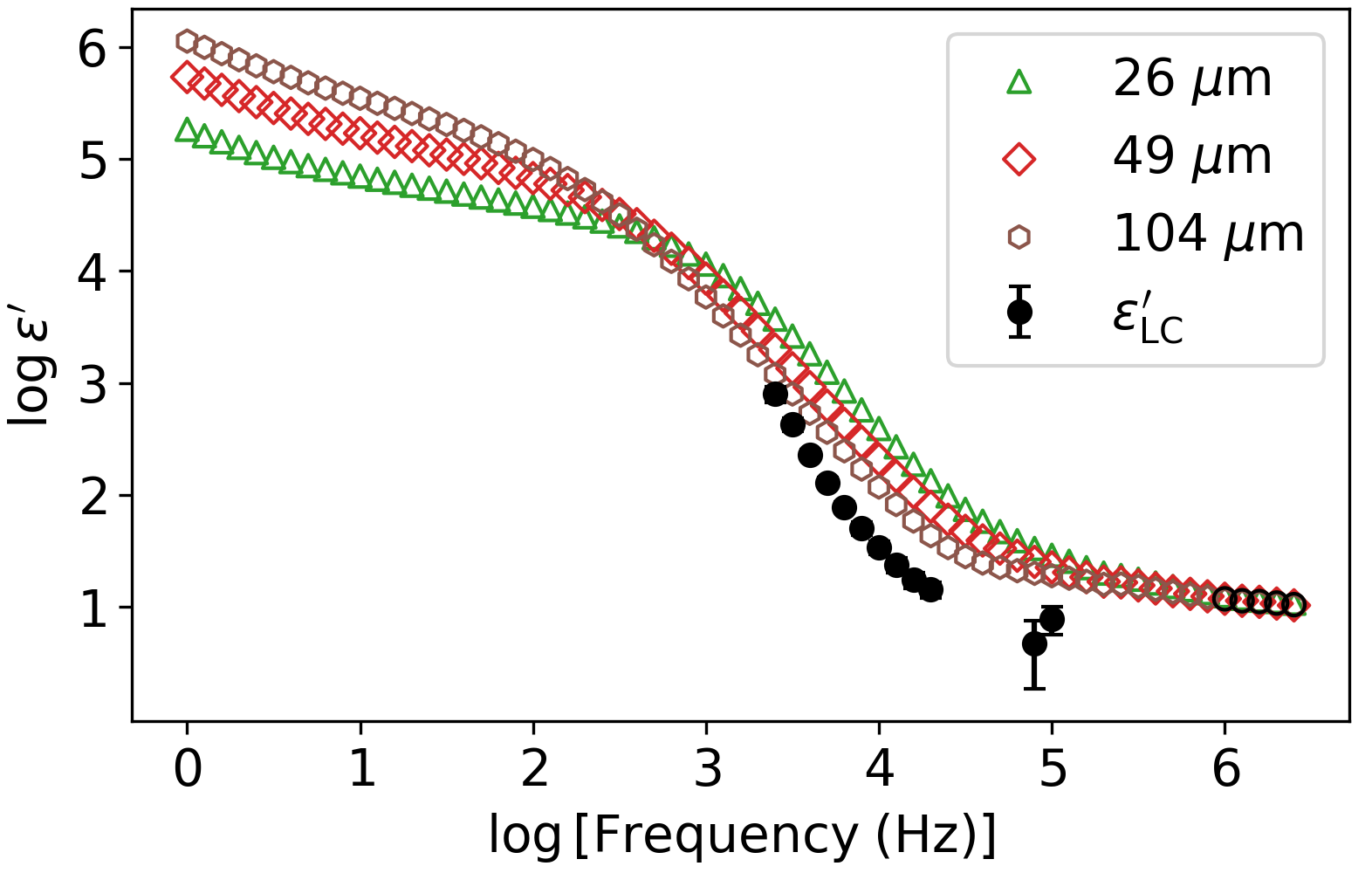}
\end{center}
\caption{\label{fig:correction_comparison} Comparison of the corrected dielectric spectrum at $125^{\circ}$C (black symbols) with the measured curves at three different thicknesses.}
\end{figure}

\renewcommand{\thefigure}{S6}
\begin{figure}[htbp]
\begin{center}
\includegraphics[width=70 mm]{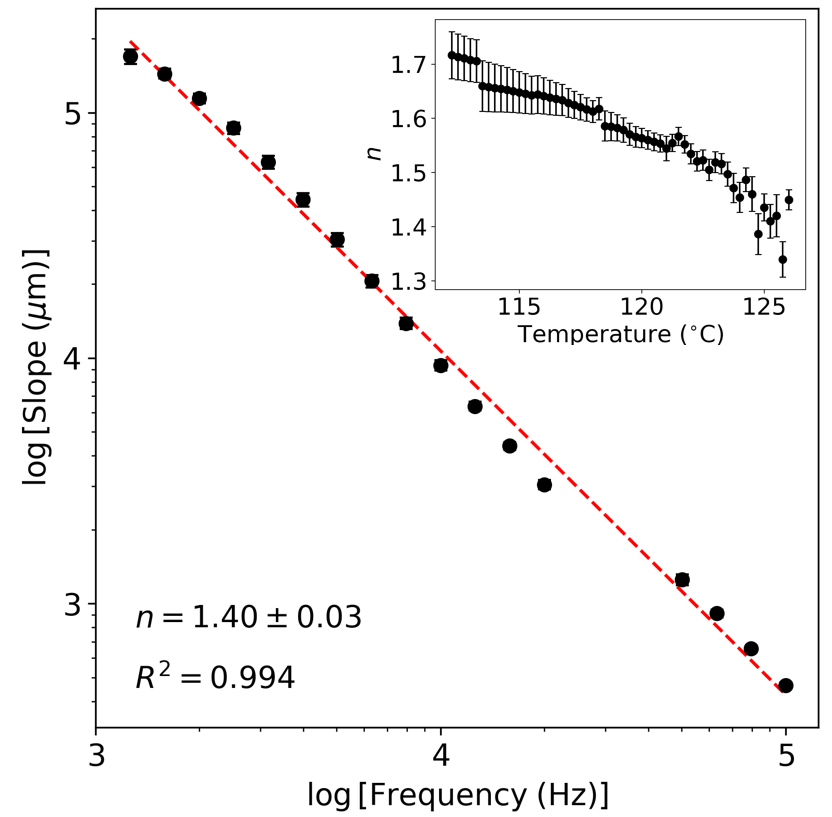}
\end{center}
\caption{\label{fig:SIexponent_exp} Frequency dependence of the slopes obtained from the fittings to the EP model in the intermediate frequency range (example in Fig. 9), where a power law is observed. Some experimental points are missing around 30 kHz because we established a $R^2\geqslant 0.99$ criterion for inclusion in the analysis. The exponent $n$ characterizing the power law can be seen to deviate from the ideal value of $2$ as evidenced by the linear regression. The inset shows that $n$ varies between $1.4$ and $1.7$ in the investigated temperature range. Some authors have suggested that this deviation is to be expected under certain conditions when indeed $n=3/2$ should be observed [34, 36, 37].}
\end{figure}




